\documentclass[pra,twocolumn,preprintnumbers,amsmath,amssymb,floatfix,superscriptaddress,showpacs]{revtex4}
\usepackage{graphicx}
\usepackage{graphics}
\usepackage{psfrag}
\usepackage{hyperref}

\newcommand{\vphi}{\varphi}

\newcommand{\vare}{\varepsilon}
\newcommand{\sgn}{\mbox{sgn}}
\newcommand{\rmi}{{\rm i}}

\begin{document}

\hypersetup{pdftitle={Tan contact and universal high momentum behavior of the fermion propagator in the BCS-BEC crossover}}
\title{Tan contact and universal high momentum behavior of the\\fermion propagator in the BCS-BEC crossover}
\date{\today}
\author{Igor Boettcher}
\affiliation{Institute for Theoretical Physics, University of Heidelberg, D-69120 Heidelberg, Germany}
\author{Sebastian Diehl}
\affiliation{Institute for Theoretical Physics, University of Innsbruck, A-6020 Innsbruck, Austria}
\affiliation{Institute for Quantum Optics and Quantum Information of the Austrian Academy of Sciences, A-6020 Innsbruck, Austria}
\author{Jan M. Pawlowski}
\affiliation{Institute for Theoretical Physics, University of Heidelberg, D-69120 Heidelberg, Germany}
\affiliation{ExtreMe Matter Institute EMMI, GSI Helmholtzzentrum f\"{u}r Schwerionenforschung mbH, D-64291 Darmstadt, Germany}
\author{Christof Wetterich}
\affiliation{Institute for Theoretical Physics, University of Heidelberg, D-69120 Heidelberg, Germany}

\begin{abstract}
We derive the universal high momentum factorization of the fermion self-energy in the BCS-BEC crossover of ultracold atoms using nonperturbative quantum field theoretical methods. This property is then employed to compute the Tan contact as a function of interaction strength, temperature, and chemical potential or density. We clarify the mechanism of the factorization from an analysis of the self-consistent Schwinger--Dyson equation for the fermion propagator, and compute the perturbative contact on the BCS and BEC sides within this framework. A Functional Renormalization Group approach is then put forward, which allows to determine the contact over the whole crossover and, in particular, for the Unitary Fermi gas. We present numerical results from an implementation of the Renormalization Group equations within a basic truncation scheme.
\end{abstract}


\maketitle


\section{Introduction}
The physics of strongly correlated many-body systems plays a key role for the understanding of nature in a wide range of energy scales, occurring in situations such as high temperature superconductors, heavy ion collisions, or neutron stars. Therefore, it is desirable  to identify generic features which arise beyond a given system under consideration, and shed light on the behavior of apparently different physical settings. The recent experimental progress in ultracold atoms research opens up the possibility to study quantum many-body systems under idealized conditions with full control over the system parameters.

A particularly promising observable in this context is the equation of state, which encodes the thermodynamic properties and the macroscopic phases of the many-body system \cite{Salomon,Navon07052010,Ku03022012}. Thermodynamic functions can be expressed in terms of momentum space integrals involving single particle distribution functions. Due to the phase space factor $V\mbox{d}^3 q/(2\pi\hbar)^3$ being proportional to $q^3$, these observables are also influenced by particles with high momentum. It is a relevant question to ask which quantitative effect the large momentum tails of the distribution functions may have on the equation of state.

The single particle momentum distribution $n_{\vec{q}}$ of a noninteracting Fermi gas
in equilibrium is described by the Fermi--Dirac distribution, which
decays exponentially for momenta sufficiently larger than the chemical
potential. In contrast, for ultracold fermions in the BCS-BEC
crossover one finds an algebraic decay \cite{HaussmannZPhys91,PhysRevB.49.12975,PhysRevA.69.013607}
 according to
\begin{align}
\label{1-1} n_{\psi,\vec{q}\sigma} = \frac{C}{q^4}\ \text{for large $q$}.
\end{align}
Here $n_{\psi,\vec{q}\sigma}$ is the occupation number of fermions of species $\sigma$ and
the precise meaning of ``large $q$'' will be specified later. It
was realized by Tan  \cite{Tan20082952,Tan20082971,Tan20082987} that the factor of proportionality $C$, called
contact, makes its appearance in several exact relations describing
the quantum many-body system. For instance, it allows to compute the
change of the energy density of the system due to a change of the
scattering length according to the adiabatic sweep theorem
\begin{align}
\label{1-2} \frac{1}{V}\frac{\mbox{d}E}{\mbox{d}(-1/a)} = \frac{C}{4\pi M}.
\end{align}
Moreover, the energy density of the homogeneous system is given by
\begin{align}
  \label{1-3} \frac{E}{V} = \frac{C}{4\pi M a} + \sum_{\sigma=1,2}
  \int \frac{\mbox{d}^3q}{(2\pi)^3} \frac{q^2}{2M}
  \Bigl(n_{\psi,\vec{q}\sigma}-\frac{C}{q^4}\Bigr).
\end{align}
Interestingly,  there also is a connection between the contact and the
shear viscosity of the system  \cite{PhysRevLett.99.170404,PhysRevA.81.053610}. The Tan relations \cite{PhysRevLett.100.205301,Braaten:2008bi,PhysRevA.79.023601,Braaten:2010if,PhysRevA.81.021601,PhysRevA.81.063634,0295-5075-91-2-20005} have found
verifications in experiments on the BCS-BEC crossover of ultracold
fermionic atoms \cite{PhysRevLett.95.020404,PhysRevLett.95.120402,CastinWerner,PhysRevLett.104.235301,PhysRevLett.105.070402,PhysRevLett.106.170402}. The measurement of the contact for a homogeneous system has recently been obtained in \cite{JinContact}.

The contact $C$ is related in a universal manner to the high momentum
behavior of the fermion propagator. To elucidate this connection we consider
the euclidean  propagator of species $\sigma=1,2$ for two atomic hyperfine states,
\begin{align}
 \label{1-4} G_{\psi \sigma}(Q) = \frac{1}{P_{\psi,\rm cl}(Q)+\Sigma_\psi(Q)} = \frac{1}{\rmi q_0+q^2-\mu+\Sigma_\psi(Q)},
\end{align}
where $Q=(q_0,\vec{q})$ denotes both euclidean frequency and momentum. We set $\hbar=k_{\rm B}=2M=1$
with fermion mass $M$ and denote $q^2=|\vec{q}|^2$. After analytic continuation $\rmi q_0 \to -\omega$, the zeros of $P_{\psi,\rm cl}(\omega_q,q^2)$ result in a classical dispersion relation $\omega_q=q^2-\mu$. For this reason we call $P_{\psi,\rm cl}(Q)=\rmi q_0+q^2-\mu$ the classical
inverse propagator.

The full fermion propagator has a nontrivial momentum dependence which goes beyond the quadratic behavior
of the classical one. This is encoded in the complex valued self-energy $\Sigma_\psi(Q)$, which we define in Eq. (\ref{1-4}), and which is a function of frequency, momentum, and system parameters like temperature, chemical potential or scattering length. It is a challenging task of quantum many-body physics to (approximately) compute the self-energy for a given system under consideration. In particular, for strongly correlated systems like the Unitary Fermi gas no obvious expansion scheme is at hand and one has to rely on nonperturbative methods like Monte Carlo simulations \cite{PhysRevLett.93.200404,PhysRevLett.95.230405,PhysRevA.81.033619,PhysRevLett.106.205302}, self-consistent T-matrix approaches \cite{HaussmannZPhys91,PhysRevB.49.12975,PhysRevA.80.063612,PhysRevA.82.021605,Enss2011770,1367-2630-13-3-035007}, $1/\mathcal{N}$-expansion \cite{PhysRevA.86.013616}, $\epsilon$-expansion \cite{PhysRevLett.97.050403}, Schwinger--Dyson equations (SDE) \cite{PhysRevA.73.033615,PhysRevA.77.023626} or the Functional Renormalization Group (FRG) \cite{Birse:2004ha,PhysRevA.76.021602,PhysRevA.76.053627,ANDP:ANDP201010458}.

Using SDE and the FRG we derive the universal factorization property of the fermion self-energy
\begin{align}
\label{1-5} \Sigma_{\psi}(Q) =\frac{4C}{P_{\psi,\rm cl}(-Q)} - \delta \mu\mbox{ }\ \text{for large $q$},
\end{align}
where $q$ has to be larger than any of the physical scales $k_{\rm ph}$ set by inverse scattering length $a^{-1}$, chemical potential $\mu$ and temperature $T$. The first contribution in Eq. (\ref{1-5}), which we refer to as Tan term, results in the large momentum decay of the momentum distribution according to $n_{\psi,\vec{q}\sigma} = C/q^4$, see Eq. (\ref{sd14}) below. The second term constitutes a high momentum shift of the effective fermion chemical potential (Hartree shift). Within SDE it is not trivial to show that $\Sigma_\psi$, $C$ and $\delta \mu$ are ultraviolet finite quantities that do not involve any counterterms. This requires a suitable split of the relevant momentum integrals, see Eq. (\ref{sd5}) below.

The importance of the double fraction structure of the fermion propagator in order to quantitatively describe the BCS-BEC crossover has first been pointed out by Haussmann in Refs. \cite{HaussmannZPhys91,PhysRevB.49.12975}.   Universal high momentum factorization of dynamic quantities has also been observed with the FRG in the context of finite temperature Yang--Mills theory \cite{Fister:2011uw} and with the Similarity Renormalization Group applied to $N$-body systems at zero temperature, including deuteron, ultracold fermions and the electron gas \cite{PhysRevC.82.054001,Bogner:2012zm}. The large momentum behavior of the self-energy is related to properties of energetic atoms propagating in a strongly interacting gas, see Ref. \cite{PhysRevA.85.053643} for a operator product expansion study.

Although the contact $C$ appears in the high momentum propagator of the theory, it is a many-body quantity dominated by interaction and many-body scales. This is also reflected in its close relation to thermodynamic quantities. Thermodynamic considerations based on the Tan relations restrict the most general form of the contact as a function of temperature \cite{PhysRevA.80.023615,1367-2630-13-3-035007}. In order to compute the function $C(\mu,T,a)$ it is therefore mandatory to work within a setting which can resolve the physics on all scales of the theory. This is realized by an FRG framework, where fluctuations on distinct scales are integrated out successively, and which goes beyond the mean-field approximation. 

A main result of this work is the development of an FRG scheme to compute the physics related to the contact. In particular, we derive a flow equation for the flowing contact $C_k$, which interpolates between $C_{k=\Lambda}=0$ in the ultraviolet and the physical contact $C_{k=0}=C$ in the infrared. The method readily applies to nonzero values of the crossover parameter $(k_{\rm F}a)^{-1}$ and any spatial dimension $d$. It allows for improving quantitative precision by using more elaborate truncations. We come back to this point in our discussion in Sec. \ref{SecDis}. For the purpose of the present work we restrict to $d=3$. Moreover, we only consider the spin-balanced case of an equal population of fermions of species $\sigma=1, 2$.

We normalize the momentum distribution such that the integral $n=2\int_{\vec{q}} n_{\vec{q}\sigma}$ yields the density of atoms $n$, and thus the contact $C$ is an intensive quantity. Defining $N=2\int_{\vec{q}} \bar{n}_{\vec{q}\sigma}$ instead, with the particle number $N$, results in an extensive contact $\bar{C}=CV$, where $V$ is the volume of the system. We have $\bar{C}/Nk_{\rm F} = 3 \pi^2 C/k_{\rm F}^4$.

An interesting question is related to the range of applicability of
the asymptotic formula (\ref{1-5}) for the fermion propagator,
i.e. the momentum scale $q_{\rm c}$ such that the universal scaling form is
valid for $q \geq q_{\rm c}$. This has direct consequences for the density
of the system. Indeed, as was already noted in \cite{Tan20082971}, the contribution
from high energetic atoms to the total atom density for $T=0$ can be
approximated by
\begin{align}
\label{1-6} \delta n^{(C)} = 2 \int_{q\geq q_{\rm c}} \frac{\mbox{d}^3q}{(2\pi)^3} \frac{C}{q^4} =  \frac{C}{\pi^2q_{\rm c}}.
\end{align}
The apparent divergence of this expression for $q_{\rm c}\to 0$ is cured by a nonzero gap or temperature, such that $\delta n^{(C)}$ remains finite. Thus, there is no a priori lower bound for $q_{\rm c}$.

It can be understood easily that $q_{\rm c}$ is smallest for the Unitary Fermi gas. Indeed, the universal form of the self-energy in Eq. (\ref{1-5}) is valid for $q$ being larger than the physical scales $k_{\rm ph}$. Typically, the inverse scattering length $a^{-1}$ is much larger than the momentum scales set by chemical potential and temperature. In particular, this is valid in the perturbative regimes on the BEC and BCS sides, where $|a|\to0$. However, for the Unitary Fermi gas we have $a^{-1}=0$ and the first physical scale is set by either $\mu^{1/2}$ or $T^{1/2}$. For this reason there is a huge \emph{enhancement of the universal regime} where Eq. (\ref{1-5}) is valid in the unitary limit. This is seen in Sec. \ref{SecFRG}. From Eq. (\ref{1-6}) we then conclude that the contribution of high energetic particles on the thermodynamic functions is large at resonance.

This paper is organized as follows. In Sec. \ref{SecMIC} we introduce the microscopic model which is employed throughout the paper. The factorization property of the fermion self-energy is deduced from the SDE for the fermion propagator in Sec. \ref{SecSDE}. We derive the expected results $C_{\rm BEC}=4\pi n/a$ and $C_{\rm BCS}=4 \pi^2a^2n^2$ for the contact in the perturbative regimes, with the detailed calculation presented in App. \ref{AppSDE}. In Sec. \ref{SecFRG} we formulate the FRG approach to the contact physics. After general remarks on the FRG we derive the factorization of the flow of the fermion self-energy at large external momenta. The result is extended to the ordered regime with a nonvanishing field expectation value and we extract a flow equation for the flowing contact. We show that the contact becomes sizeable at the many-body scales of the system. In Sec. \ref{SecRes} we estimate the range of applicability of the universal scaling behavior for the Unitary Fermi gas, compare the zero temperature result to perturbation theory and determine the temperature dependence of the contact at resonance. We draw our conclusions and discuss the influence of the contact on the particle density and the critical temperature $T_{\rm c}/T_{\rm F}$ in Sec. \ref{SecDis}. In App. \ref{AppAtoms} we relate our discussion of the two-channel model to the contact in a purely fermionic language. In Apps. \ref{AppA} and \ref{AppB} we present the basic truncation which is employed here and derive the flow equation for the flowing contact. We recapitulate the derivation of the contact on the BEC and BCS sides from the perturbative equation of state at zero temperature in App. \ref{AppC}.

\section{Microscopic model}
\label{SecMIC}

We consider a system of ultracold two-component fermions described by the microscopic many-body Hamiltonian
\begin{align}
\label{2-1} \hat{H} = \int \mbox{d}^3x \biggl( \sum_{\sigma=1,2} \hat{\psi}^\dagger_\sigma (-\nabla^2) \hat{\psi}_\sigma + \lambda_{\psi,\Lambda} \hat{\psi}^\dagger_1\hat{\psi}^\dagger_2\hat{\psi}_2\hat{\psi}_1\biggr).
\end{align}
Here $\hat{\psi}^{(\dagger)}_\sigma(\vec{x})$ is the annihilation (creation) operator of a fermion atom in hyperfine state $\sigma$ at position $\vec{x}$.  Scattering between atoms is modeled by a local contact interaction with coupling constant $\lambda_{\psi,\Lambda}<0$, which eventually translates into an s-wave scattering length $a$. This effective description provides a solid starting point to compute observables for experiments with dilute ultracold alkali atoms such as $^6$Li or $^{40}$K, since therein small distance details of the interatomic potential need not be resolved. We make this explicit by defining the theory with reference to a highest momentum scale $\Lambda$. Physics on distances smaller than $\Lambda^{-1}$ cannot be deduced within the pointlike approximation and requires more information on the atomic interactions, like the effective range $r_{\rm e}$ or the van der Waals length $\ell_{\rm vdW}$. We choose $\Lambda$ sufficiently large but ensure $\Lambda^{-1} \gg r_{\rm e}, \ell_{\rm vdW}$. In the spirit of this consideration we refine our definition of the contact according to
\begin{align}
\label{2-2} C = \lim_{\stackrel{q\to\infty,}{q\ll \Lambda}} q^4 n_{\vec{q}\sigma}.
\end{align}

Our analysis builds on a field theoretical formulation using functional integral techniques. For this purpose we define the microscopic model of Eq. (\ref{2-1}) in terms of the euclidean action
\begin{align}
\label{2-3} S = \int_X \biggl(\sum_{\sigma=1,2}\psi^*_\sigma(\partial_\tau-\nabla^2-\mu)\psi_\sigma + \lambda_{\psi,\Lambda} \psi^*_1\psi^*_2\psi_2\psi_1\biggr).
\end{align}
The Grassmann valued field $\psi_\sigma(\tau,\vec{x})$  depends on the euclidean time variable $\tau$, which is restricted to a torus of circumference $\beta=1/T$. Adjusting $T$ and the chemical potential $\mu$ allows for a description of the theory both at zero and nonzero temperature and density in a unified framework. We employ the abbreviation 
\begin{align}
 \label{2-4} \int_X=\int_0^\beta\mbox{d}\tau\int\mbox{d}^3x.
\end{align}

We work with a partially bosonized version of the action (\ref{2-3}), where the contact interaction of fermion atoms is modeled by exchange of a bosonic di-atom state. The bosonic field enters the microscopic action by virtue of a Hubbard--Stratonovich transformation, which yields
\begin{align}
 \nonumber S_\Lambda[\vphi,\psi] &=\int_X \biggl( \sum_{\sigma=1,2}\psi^*_\sigma(\partial_\tau -\nabla^2-\mu)\psi_\sigma \\
 \label{2-5}&+(\nu+\delta \nu_\Lambda)\vphi^*\vphi-h_\vphi(\vphi^*\psi_1\psi_2-\vphi\psi_1^*\psi_2^*)\biggr)
\end{align}
with Feshbach coupling $h_\vphi$ between fermions and bosons, and detuning $\nu=8 \pi a$. We introduce a counter\-term $\delta \nu_\Lambda$ for the constant part of the boson propagator. We observe that the four-fermion interaction has dropped out, which requires $\lambda_{\psi,\Lambda}=-h^2_{\vphi,\Lambda}/\nu_{\Lambda}$ with $\nu_\Lambda=\nu+\delta\nu_\Lambda$. Indeed, performing the Gaussian functional integral over the bosons one recovers a fermionic functional integral with action (\ref{2-3}). Whereas the four-fermion coupling gets strongly affected by fluctuations, eventually resulting in a divergence at the critical temperature, the Feshbach coupling can be approximated by its classical, momentum independent value. The scale dependence of $\lambda_\psi$ resides dominantly in the scale dependence of the term $\nu$ quadratic in the bosons.

The boson field $\vphi(\tau,\vec{x})$ is not a dynamical degree of freedom in this microscopic action which is thus equivalent to Eq. (\ref{2-3}). However, a dynamical part of the boson propagator will be generated due to quantum and thermal fluctuations. On the BCS side of the crossover $\vphi$ describes Cooper pairs of atoms, whereas it refers to bosonic molecules on the BEC side. Bose condensation of pairs, i.e. a nonvanishing expectation value of the fluctuating field $\vphi$, leads to superfluidity of the system. We will find below a connection between the contact $C$ and the number of atoms bound in (not necessarily condensed) bosonic pairs, in line with Tan's interpretation  based on  the many-body Schroedinger equation \cite{Tan20082952,Tan20082971,Tan20082987}.

\section{Schwinger--Dyson equations}
\label{SecSDE}
The physics of the contact is based on a separation of scales. This scale dependence is most efficiently resolved within a Renormalization Group analysis. Nevertheless it is instructive to first consider a self-consistent gap equation or Schwinger--Dyson equation for the fermion propagator. From this exact equation, the factorization property of the self-energy at large external momenta is deduced easily. Moreover, we use this formalism to show how the perturbative results $C_{\rm BEC}=4\pi n/a$ and $C_{\rm BCS}=4 \pi^2a^2n^2$ arise naturally on the BEC and BCS sides of the crossover after a proper ultraviolet renormalization scheme has been applied.

From Eq. (\ref{2-5}) we infer the inverse classical fermion and boson propagators, respectively, to be given by
\begin{align}
 \nonumber P_{\psi, \rm cl}(q_0,q^2) &= \rmi q_0 +q^2 -\mu,\\
 \label{sd2} P_{\vphi,\rm cl}(q_0,q^2) &= \nu +\rmi \vare q_0.
\end{align}
We introduce an infinitesimal contribution $\rmi \vare q_0$ with $\vare \to 0^+$ in the inverse boson propagator to regularize momentum integrals involving $P_\vphi$. With this single modification we will make manifest that only one single coupling, the detuning $\nu$, requires an ultraviolet renormalization in the BCS-BEC crossover, see Eqs. (\ref{sd20}) and (\ref{sd21}). Alternatively, one can add a counterterm $\propto f_\Lambda \psi^*_\sigma\psi_\sigma$ in the fermionic part of the microscopic action and then adjust $f_\Lambda$ appropriately.  At nonzero temperature, the variable $q_0$ is discretized according to the Matsubara formalism to $q_0=2\pi (n+1/2)T$ ($2\pi n T$) for fermions (bosons) with $n\in\mathbb{Z}$. We write
\begin{align}
 \label{sd3} \int_Q = T \sum_n \int \frac{\mbox{d}^3q}{(2\pi)^3}.
\end{align}
For $T=0$ the sum is replaced by the corresponding Riemann integral.

\begin{figure}[t!]
 \centering
 \includegraphics[bb=0 0 293 71,scale=0.8,keepaspectratio=true]{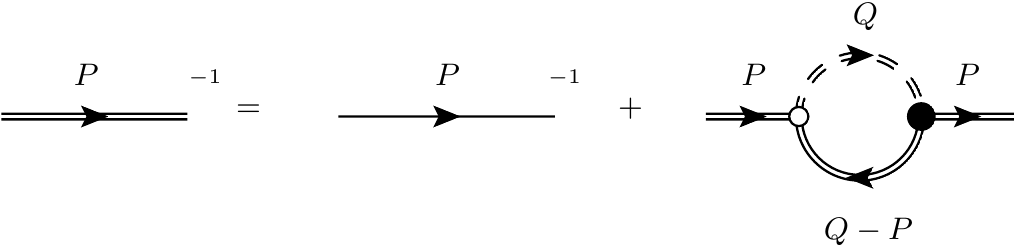}
 \caption{Schwinger--Dyson equation for the inverse fermion propagator. A single line corresponds to a classical propagator and a double line denotes a full propagator. Fermions and bosons are represented by solid and dashed lines, respectively. In the loop integral we have one fully dressed Feshbach coupling and a microscopic one. The latter is momentum independent.}
 \label{FigFermSDE}
\end{figure}

Due to quantum and thermal fluctuations, the classical propagators $P_{\psi,\rm cl}(Q)$ and $P_{\vphi, \rm cl}(Q)$ get dressed to yield the macroscopic propagators $P_\psi(Q)$ and $P_\vphi(Q)$. This correction is encoded in the  self-energies $\Sigma_{\psi/\vphi}(Q) = P_{\psi/\vphi}(Q)-P_{\psi/\vphi,\rm cl}(Q)$. The Schwinger--Dyson equation for the full inverse fermion propagator reads
\begin{align}
 \label{sd4} P_\psi(Q) = P_{\psi,\rm cl}(Q) + \int_P \frac{h_\vphi^2}{P_\psi(P-Q)P_\vphi(P)}
\end{align}
according to the graph shown in Fig. \ref{FigFermSDE}. The corresponding momentum integral involves the fully dressed propagators. It is in this sense that the equation is self-consistent and cannot be solved in a straightforward manner. In Eq. (\ref{sd4}) we approximate the full Feshbach coupling by the microscopic one, which is momentum independent. This approximation is justified due to the weak effect of fluctuations on the Feshbach coupling.

The BCS-BEC crossover across a broad Feshbach resonance can be parametrized by the physical scales temperature, chemical potential and scattering length. We denote the highest physical momentum scale by $k_{\rm ph}$. In the perturbative regimes, this corresponds to the inverse scattering length, whereas this is not valid close to resonance where $a^{-1}=0$. Only momenta $p^2 < k^2_{\rm ph}$ in the loop-integral in Eq. (\ref{sd4}) can resolve the details of the choice of parameters. For this reason we split up the integration according to
\begin{align}
 \nonumber \Sigma_\psi(Q) = &\mbox{ }h_\vphi^2T\sum_n\int_{p^2< k_{\rm tr}^2}  \frac{1}{P_\psi(P-Q)P_\vphi(P)} \\
 \label{sd5} &+ h_\vphi^2T\sum_n\int_{p^2 > k_{\rm tr}^2} \frac{1}{P_\psi(P-Q)P_\vphi(P)},
\end{align}
where the transition momentum $k_{\rm tr}$ is sufficiently larger than $k_{\rm ph}$, such that for $p^2 > k_{\rm tr}^2$ the $p^2$-dependence of $P_\vphi(P)$ can be neglected. We can then replace 
\begin{align}
 \label{sd5b} P_\vphi(P) \to P_\vphi(p_0,p^2=k^2_{\rm tr}) =: P_{\vphi, \rm tr}(p_0) 
\end{align}
in the second integral. By adding and subtracting a convenient piece we can now write $\Sigma_\psi(Q)$ in terms of two contributions, $\Sigma_\psi = \Sigma_\psi^{(1)}+\Sigma_\psi^{(2)}$, with
\begin{align}
 \nonumber &\Sigma_\psi^{(1)}(Q) \\
 \nonumber &= h_\vphi^2 T \sum_n \int_{p^2< k^2_{\rm tr}} \frac{1}{P_\psi(P-Q)}\biggl(\frac{1}{P_\vphi(P)}-\frac{1}{P_{\vphi, \rm tr}(p_0)}\biggr),\\
 \label{sd5c} &\Sigma_\psi^{(2)}(Q) = h_\vphi^2 \int_P \frac{1}{P_\psi(P-Q)P_{\vphi, \rm tr}(p_0)}.
\end{align}
Both pieces are manifestly ultraviolet finite and do not depend on $k_{\rm tr}$ for sufficiently large $k^2_{\rm tr} \gg k^2_{\rm ph}$.

We emphasize that the splitting of the self-energy in Eq. (\ref{sd5}) enables us to show that the superficially divergent loop-integral in Eq. (\ref{sd4}) is indeed finite for large $q^2$. This nontrivial statement results from a scale argument only and thus does not imply any restrictions on coupling strength, density, or temperature. Moreover, the particular choice of the artificially introduced momentum $k_{\rm tr}$ does not play a role for the final result.

We are interested in the behavior of $\Sigma_\psi(Q)$ for large momenta $q^2 \gg k_{\rm tr}^2$. The scale hierarchy in this case is given by
\begin{align}\label{hierarchy}
 k_{\rm ph}^2 \ll k_{\rm tr}^2 \ll q^2 \ll \Lambda^2.
\end{align}
In $\Sigma_\psi^{(1)}(Q)$ we can then replace $P_\psi(P-Q)$ by $P_{\psi,\rm cl}(-Q)$, because the integral is restricted to momenta which are small in comparison to $Q$ and we can neglected self-energy corrections to leading order  at high momenta. This results in the factorization
\begin{align}
 \label{sd5d} \Sigma_\psi^{(1)}(Q) = \frac{4C}{P_{\psi,\rm cl}(-Q)}\ \text{for large }q^2.
\end{align}
We define the contact according to
\begin{align}
 \label{sd5e} C = \frac{h_\vphi^2}{4} T\sum_n \int_{p^2 < k^2_{\rm tr}} \biggl(\frac{1}{P_\vphi(P)}-\frac{1}{P_{\vphi, \rm tr}(p_0)}\biggr),
\end{align}
since the numerator in Eq. (\ref{sd5d}) is seen below to result in the prefactor $C$ of the $1/q^4$-tail of the momentum distribution. Again, formulas (\ref{sd5d}) and  (\ref{sd5e}) only rely on the splitting of the loop-integral at $k_{\rm tr}$, which gives us a precise notion of ``large $q^2$'', namely $q^2 \gg k_{\rm tr}^2$. The factorization is thus a generic feature of theories which can be described by similar Feshbach- or Yukawa-type gap equations.

We can define an effective boson occupation number as
\begin{align}
 \label{sd5f} n_{\vphi,\vec{p}} = T \sum_n \biggl(\frac{1}{P_\vphi(P)}-\frac{1}{P_{\vphi, \rm tr}(p_0)}\biggr)
\end{align}
with total boson number density $n_\vphi = \int_{\vec{p}} n_{\vphi,\vec{p}}$. This yields
\begin{align}
 \label{sd5g} C = \frac{h_\vphi^2}{4} n_\vphi
\end{align}
and hence shows the close relation between the contact and the number of atoms bound in bosonic pairs. Note that the definition of $n_{\vphi}$ has to be renormalization group invariant and thus involves a wave function renormalization constant, which we set to unity here for simplicity. The second term in $\Sigma_\psi(Q)$ results in a shift of the effective chemical potential
\begin{align}
 \label{sd5h} \Sigma_\psi^{(2)}(Q) = -\delta\mu\ \text{for large }q^2.
\end{align}

The correction to the momentum distribution of particles at high momenta which results from the self-energy in Eqs. (\ref{sd5d}) and (\ref{sd5h}) is found from the generally valid formula
\begin{align}
 \label{sd9}n_{\psi,\vec{q}\sigma} = - \biggl( T \sum_n \frac{1}{P_\psi(Q)} -\frac{1}{2} \biggr).
\end{align}
Treating the self-energy perturbatively for large external momentum $q^2$, we find
\begin{align}
 \nonumber n_{\psi,\vec{q}\sigma}= &-\biggl( T \sum_n \frac{1}{P_{\psi,\rm cl}(Q)-\delta\mu}-\frac{1}{2}\biggr)\\
  \label{sd10}&+ T \sum_n \frac{4C}{P_{\psi,\rm cl}^2(Q)P_{\psi,\rm cl}(-Q)}.
\end{align}
The first two terms yield $N_F(q^2-\mu-\delta\mu)$ with Fermi function $N_F(z)=(e^{z/T}+1)^{-1}$. This contribution vanishes for large $q^2$. Evaluating the Matsubara summation we find for the second contribution
\begin{align}
 \nonumber n_{\psi,\vec{q}\sigma} &= T \sum_n \frac{4C}{P_{\psi,\rm cl}^2(Q)P_{\psi,\rm cl}(-Q)}\\
 \nonumber &= 4C \Bigl(\frac{1-N_F(q^2-\mu)}{4(q^2-\mu)^2}+\frac{N_F'(q^2-\mu)}{2(q^2-\mu)}\Bigr)\\
 \label{sd14} &\stackrel{q^2\gg \mu,T}{\longrightarrow} \frac{C}{q^4}\ \text{for large } q^2.
\end{align}
This justifies the identification of $C$ in the numerator of the asymptotic self-energy (\ref{sd5d}) with the contact as defined by Eq. (\ref{2-2}). Higher order contributions to the fermion self-energy do not enter the $1/q^4$-tail of the momentum distribution.

The formulas derived in this section become particularly simple in the perturbative BEC and BCS regimes, because the integrals in Eq. (\ref{sd5c}) can be performed analytically. We present the calculation in App. \ref{AppSDE} and give here only the results. The shift $\delta \mu$ of the effective chemical potential vanishes on the BEC side of the crossover. On the BCS side, it is given by
\begin{align}
 \delta \mu = -\lambda_{\psi} n_{\psi,\sigma},
\end{align}
with four-fermion coupling $\lambda_\psi=8 \pi a$. Since all atoms are bound to dimers in the BEC limit, the boson density in Eq. (\ref{sd5g}) equals half the particle density and we arrive at
\begin{align}
 C_{\rm BEC} = \frac{4 \pi n}{a}.
\end{align}
The relation $h^2_\vphi=32 \pi /a$ results from the wave function renormalization of the boson propagator, see Eq. (\ref{sd22}). In the BCS limit, the first nonvanishing contribution to the contact arises at second order in perturbation theory in the coupling $a$. Inserting the Schwinger--Dyson equation for the boson propagator (\ref{sd20}) into Eq. (\ref{sd5g}), we obtain a double integral over two fermion propagators, each resulting in a fermion density $n_{\psi,\sigma}$. We find
\begin{align}
 C_{\rm BCS} = 4 \pi^2 a^2 n^2.
\end{align}
These perturbative results derived from Schwinger--Dyson equations agree with the expressions found from the zero temperature equation of state. The corresponding calculation is recapitulated in App. \ref{AppC}.

\section{Functional Renormalization Group}
\label{SecFRG}

The factorization (\ref{sd5d}) relies on the scale hierarchy
(\ref{hierarchy}). The latter is also at the root of the derivation of the
contact $C$ in Eq. (\ref{sd5e}) by means of the splitting of the integration defined
in Eq. (\ref{sd5}). A method of choice which naturally incorporates such a
momentum splitting and the factorization in the fermionic self-energy is
the Functional Renormalization Group (FRG) approach to the BCS-BEC
crossover. This has been already shown in the FRG approach to thermal
propagators in Yang--Mills theory \cite{Fister:2011uw}. 

We will use the flow equation for the effective average action \cite{Wetterich199390}. It has a one-loop form for which momentum integrals are both ultraviolet and infrared finite, dominated by a small momentum range in the vicinity of the infrared cutoff, $q^2 \approx k^2$. As a direct consequence, the splitting into different momentum ranges as in Eq. (\ref{sd5}) can be avoided, and no particular care has to be taken for the construction of finite quantities. This constitutes an important advantage as compared to the Schwinger--Dyson equations, especially in the region of unitarity, where the validity of Eq. (\ref{1-5}) extends to rather small momenta.

Thus use of FRG allows us to extract the many-body quantity $C(a,\mu,T)$ in
the whole crossover 
with $k_{\rm tr}$. It also goes beyond the universal scaling
behavior proposed in Eq. (\ref{1-5}). In fact, the only assumption
underlying the following analysis is the \emph{perturbative} treatment
of the fermion propagator at high momenta, whereas the bosonic degrees
of freedom are treated in a nonperturbative fashion. The latter is in
particular mandatory to resolve the second order nature of the
superfluid phase transition.

\subsection{Flow equation}

We start from the microscopic action of the two-channel model given in Eq. (\ref{2-5}) at a high momentum scale $\Lambda$ and successively include quantum and thermal fluctuations on momentum scales larger than an infrared cutoff $k$. By lowering $k$ we resolve the macroscopic properties of the system in a coarse graining procedure and, eventually, arrive at the full effective action $\Gamma$ of the theory for $k=0$. The effective action generates all correlation functions of the system. (More precisely, it is the generating functional of the one-particle irreducible Green functions, which are obtained by simple functional differentiation.) In between, the effective average action $\Gamma_k$ can be interpreted as the effective action for a probe with size $k^{-1}$. Although $\Gamma_k$ for $k>0$ is not a physical observable, as it depends on the scheme in which we lower $k$, this rough picture provides an intuitive and often surprisingly accurate guiding principle. In summary, the effective average action satisfies
\begin{align}
 \label{3-1} \Gamma_{k=\Lambda} = S,\ \Gamma_{k=0} = \Gamma.
\end{align}

The evolution of $\Gamma_k$ as $k$ is lowered is governed by the exact flow equation \cite{Wetterich199390}
\begin{align}
 \label{3-3} \partial_k \Gamma_k[\phi,\psi] = \frac{1}{2} \mbox{STr} \biggl( \frac{1}{\Gamma^{(2)}_k[\phi,\psi]+R_k}\partial_k R_k\biggr),
\end{align}
where STr is the ``supertrace'', which takes into account the mixed bosonic and fermionic structure of the theory, and $R_k$ is the regulator related to the mentioned way how $k$ is lowered. In a momentum representation, the kernel $R_k(Q)\delta(Q+Q')$ of the operator $R_k$ needs to satisfy
\begin{align}
 \label{3-3R1} \lim_{q^2/k^2\to 0} R_k(Q) \sim k^2,\ \lim_{q^2/k^2\to\infty} R_k(Q) =0.
\end{align}

If we were able to solve the functional differential equation (\ref{3-3}), which is equivalent to diagonalizing the full many-body Hamiltonian or computing the complete path integral, we would have solved the theory. In  most practical cases, however, approximative solutions are needed. The actual strength of the FRG is that it can deal with complex systems, both perturbatively and nonperturbatively, since the effective average action $\Gamma_k$ provides an intuitive way of organizing the inclusion of all fluctuations in the path integral and is thus particularly well suited for physically motivated approximations: We restrict the space of possible functionals $\Gamma_k$ to a limited set of  correlation functions, project the flow equation (\ref{3-3}) onto this parameter set, and solve the remaining set of equations numerically. Due to such a truncation we practically never arrive at the full effective action $\Gamma$ which generates all correlation functions, but obtain a quantity which is usually predictive enough to describe, for instance, the low-energy physics and thermodynamics of the system.  The truncations employed in this paper are described in Apps. \ref{AppA} and \ref{AppB}.

The application of the FRG to a broad variety of topics, ranging from high energy physics to condensed matter, are reviewed e.g. in \cite{Berges:2000ew,Pawlowski20072831,Gies:2006wv,Schaefer:2006sr,Delamotte:2007pf,Metzner:2011cw,Braun:2011pp,vonSmekal:2012vx}. The FRG approach to many-body physics with ultracold atoms is reviewed in \cite{Scherer:2010sv} and the lecture notes \cite{Boettcher:2012cm}.

\subsection{Fermion self-energy and contact term in the symmetric regime}
\label{SecFRGSYM}

For conceptual clarity we first isolate the contact term and the shift of the chemical potential from the flow equation of the self-energy $\Sigma_\psi(Q)=P_{\psi}(Q)-P_{\psi,\Lambda}(Q)$ in the symmetric or disordered regime, where the field expectation value of the boson field is zero. The procedure will be extended below to the ordered regime of the flow with a nonvanishing expectation value $\phi_0$. In general the self-energy is a $4 \times 4$--matrix. For equal population of the two hyperfine states $\sigma=1,2$, the most general form of the self-energy can be parametrized by two complex functions $\Sigma_\psi(P)$ and $\Sigma_{\psi,\rm an}(P)$, where the second one is called the anomalous contribution. For simplicity we neglect the anomalous self-energy in the following.  We show in App. \ref{AppA} that it does not contribute to the contact.

The flow of the inverse fermion propagator, which is identical to the flow of the self-energy, is given in the symmetric regime by
\begin{align}
 \nonumber &\partial_k \Sigma_{\psi,k}(P) =-\bar{h}^2_\phi \int_Q \\
 \nonumber &\biggl\{\frac{\partial_kR_{\phi,k}(Q)}{(\bar{P}_{\phi,k}(Q)+R_{\phi,k}(Q))^2}\frac{1}{P_{\psi,k}(Q-P)+R_{\psi,k}(Q-P)}\\
 \label{3-7} &+ \frac{\partial_kR_{\psi,k}(Q)}{(P_{\psi,k}(Q)+R_{\psi,k}(Q))^2}\frac{1}{\bar{P}_{\phi,k}(Q+P)+R_{\phi,k}(Q+P)}\biggr\}.
\end{align}
We emphasize that the external momentum $P$ is a free parameter and for each $P$ we have an individual flow equation. The Feshbach or Yukawa coupling $\bar{h}_\phi$ does not depend on momentum in our truncation. We also neglect a possible scale dependence of $\bar{h}_\phi$. The expressions $P_{\psi,k}(Q)=P_{\psi,\Lambda}(Q)+\Sigma_{\psi,k}(Q)$ and $\bar{P}_{\phi,k}(Q)$ are the full inverse propagators at scale $k$. We have
\begin{align}
\label{3-7b} P_{\psi,\Lambda}(Q) = \rmi q_0 + q^2-\mu.
\end{align}

Since the boson propagator is gapped for $k \gg k_{\rm ph}$, where $k_{\rm ph}$ is a physical scale given by either the inverse scattering length, temperature or chemical potential, we effectively only have a nonvanishing contribution to Eq. (\ref{3-7}) for $k \lesssim k_{\rm ph}$. Choosing a large external momentum $p^2 \gg k_{\rm ph}^2 \geq k^2$, we can use the property (\ref{3-3R1}) of the regulator functions $R_k(Q)$ to approximate
\begin{align}
 \label{3-10} P_k(Q\pm P) + R_k(Q \pm P) \simeq P_k(\pm P).
\end{align}
Thus, for large external momentum $p^2$ the flow of the self-energy simplifies according to
\begin{align}
 \nonumber \partial_k \Sigma_{\psi,k}(P) \simeq &- \frac{\bar{h}^2_\phi}{P_{\psi,k}(-P)}\int_Q \frac{\partial_kR_{\phi,k}(Q)}{(\bar{P}_{\phi,k}(Q)+R_{\phi,k}(Q))^2}\\
 \label{3-11} &- \frac{\bar{h}^2_\phi}{\bar{P}_{\phi,k}(P)}\int_Q \frac{\partial_kR_{\psi,k}(Q)}{(P_{\psi,k}(Q)+R_{\psi,k}(Q))^2}.
\end{align}

We now show that this results in an asymptotic self-energy $\Sigma_\psi = \Sigma_{\psi,k=0}$ of the form
\begin{align}
 \label{3-Sig11} \Sigma_{\psi}(P)  \simeq \frac{4C}{P_{\psi,\Lambda}(-P)} -\delta \mu.
\end{align}
The first term in Eq. (\ref{3-11}) yields the contact term. Indeed, for large external momenta $P$, renormalization effects on the fermion propagator are small and we can approximate $P_{\psi,k}(-P) \simeq P_{\psi,\Lambda}(-P)$. We then find the $P$-dependent term being multiplied by an integral which  receives contributions from the physical scales $k_{\rm ph}$. Since $P_{\psi,\Lambda}(-P)$ is $k$-independent, we can integrate the first term in Eq. (\ref{3-11}) and identify the contact as being given by
\begin{align}
 \nonumber 4 C &= - \int_\Lambda^0 \mbox{d}k \bar{h}^2_\phi \int_Q\frac{\partial_kR_{\phi,k}(Q)}{(\bar{P}_{\phi,k}(Q)+R_{\phi,k}(Q))^2}\\
 \label{3-14} &= \int_\Lambda^0 \mbox{d}k \tilde{\partial}_k \int_Q\frac{\bar{h}_\phi^2}{\bar{P}_{\phi,k}(Q)+R_{\phi,k}(Q)}.
\end{align}
In the second line, we introduced the formal derivative $\tilde{\partial}_k$, which only acts on the $k$-dependence of the regulator $R_k$. The advantage of this rewriting is to make the simple one-loop structure of the equations manifest. Eq. (\ref{3-14}) allows us to define a scale dependent contact $C_k$ according to the flow equation
\begin{align}
\label{3-14b} \partial_k C_k = \frac{\bar{h}_\phi^2}{4} \tilde{\partial}_k \int_Q\frac{1}{\bar{P}_{\phi,k}(Q)+R_{\phi,k}(Q)}
\end{align}
with $C_\Lambda=0$ and $C_{k=0}=C$. This flow equation and its generalization to the ordered regime are the basis for our numerical evaluation of $C$ in Sec. \ref{SecRes}.

The second term in Eq. (\ref{3-11}) contains the boson propagator evaluated for large momentum. Since the microscopic boson propagator is constant and the momentum dependence only builds up due to the renormalization group flow, this contribution to the fermion self-energy is independent of $P$ and constitutes a shift of the effective chemical potential. The asymptotic value is then equal to the one evaluated for a large momentum $k_{\rm tr}$. We conclude
\begin{align}
 \label{3-15} \delta \mu &= - \int_\Lambda^0 \mbox{d}k \frac{\bar{h}^2_\phi}{\bar{m}^2_\phi}\tilde{\partial}_k \int_Q \frac{1}{P_{\psi,k}(Q)+R_{\psi,k}(Q)}.
\end{align}
With the effective four-fermion vertex $-\bar{h}^2_\phi/\bar{m}^2_\phi =\lambda_\psi$ we find $\delta \mu =-\lambda_{\psi}n_{\psi\sigma}$.

\subsection{Contact term in the ordered regime}
For low enough temperatures, a nonvanishing expectation value $\rho_{0,k}$ of the boson field $\rho=\phi^*\phi$ appears during the renormalization group flow. If $\rho_{0,k=0}=\rho_0 >0$, we say that the system is in its superfluid phase and some of the bosons have condensed. Note that, due to interactions, the condensate fraction does not coincide with the superfluid fraction. Above the critical temperature there is a region where a nonvanishing value of $\rho_{0,k}$ appears for $k>0$ during the flow, but does not persist for $k \to 0$. We then arrive in the symmetric (normal) phase of the system. We may call this intermediate region the precondensation regime. It is characterized by local but not global superfluid order.

\begin{figure}[t!]
 \centering
 \includegraphics[bb=0 0 246 176,scale=0.99,keepaspectratio=true]{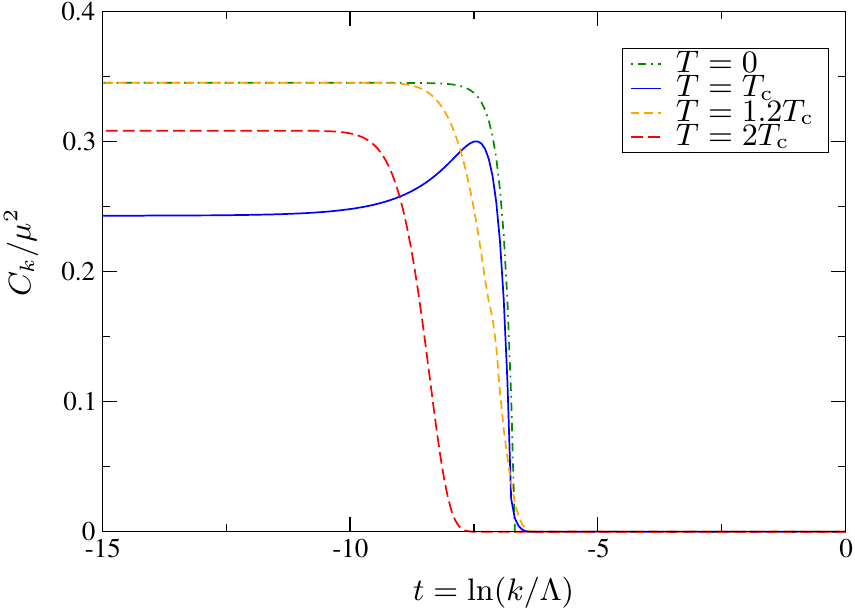}
 \caption{(Color online) RG-scale dependence of the flowing contact $C_k$ at unitarity $a^{-1}=0$. We have $k=\Lambda e^{t}$ such that $t=0$ corresponds to the ultraviolet and $t \to -\infty$ to the infrared. We observe that the contact is unaffected by fluctuations of ultraviolet modes and it starts to build up on the many-body scales of the system, which are set here by the chemical potential and temperature corresponding to $t_\mu = \ln(\mu^{1/2}/\Lambda)=-6.9$ and $t_T\simeq t_\mu$. Obviously, all curves saturate at a certain value of $t$ and  we can read off the physical value at $k=0$.}
 \label{FigFlow}
\end{figure}

Conceptually the above derivation of the flow of the self-energy and the asymptotic scaling with contact term $\sim 4C/P_{\psi,\Lambda}(-P)$ remains valid also in the presence of a possibly nonvanishing boson field expectation value. The corresponding calculation is presented in App. \ref{AppB}. We find from Eq. (\ref{B-14}) that the flow equation for the contact is given by
\begin{align}
 \label{3-29} \partial_t C_k =\mbox{ }& \frac{h^2_\phi}{4} (\eta_A \rho_{0,k}+\partial_t \rho_{0,k})\\
 \nonumber &+\frac{2^{d/2+1}v_d }{d} \frac{h_\phi^2}{S^2}k^{d+2}\Bigl(1-\frac{\eta}{d+2}\Bigr)\biggl[N_B'(\omega_{\phi,k})\\
 \nonumber &-\frac{\lambda^2\rho_0^2}{S^2\omega_{\phi,k}^3}\Bigl(\frac{1}{2}+N_B(\omega_{\phi,k})-\omega_{\phi,k}N_B'(\omega_{\phi,k})\Bigr)\biggr]
\end{align}
with
\begin{align}
 \label{3-30} \omega_{\phi,k} = \frac{\sqrt{(k^2+m^2_{\phi,k})(k^2+m^2_{\phi,k}+2\lambda_k\rho_{0,k})}}{S_k},
\end{align}
$t=\ln(k/\Lambda)$, $v_d=(2^{d+1}\pi^{d/2}\Gamma(d/2))^{-1}$ and $d=3$. The Bose function is defined as $N_B(z)=(e^{z/T}-1)^{-1}$ and $N_B'=\mbox{d}N_B/\mbox{d}z$. For a definition of the corresponding running couplings we refer to App. \ref{AppA}. Typical renormalization group flows of $C_k$ for the unitary Fermi gas are shown in Fig. \ref{FigFlow}.

At zero temperature we obtain a nonvanishing value for the contact $C$. The corresponding value is found from Eq. (\ref{3-29}) by setting the Bose functions to zero. We have
\begin{align}
 \nonumber \partial_t C_k|_{T=0} &= \frac{h_\phi^2}{4} (\eta_A \rho_{0,k}+\partial_t \rho_{0,k})\\
 \label{3-34}&-\frac{2^{d/2}v_d }{d} \frac{h_\phi^2\lambda^2\rho_0^2}{S^4\omega_{\phi,k}^3}k^{d+2}\Bigl(1-\frac{\eta}{d+2}\Bigr).
\end{align}

In the limit where the density is dominated by the superfluid density of condensed bosons, the first term in Eq. (\ref{3-29}) dominates. For small anomalous dimension $\eta_A$ and neglecting the running of $h^2_\phi$ this yields the simple relation
\begin{align}
 \label{SimRel} C \approx \frac{h^2_\phi}{4} \rho_0,
\end{align}
which coincides with Eq. (\ref{sd5g}) for $n_\vphi \approx \rho_0$.

\section{Results}
\label{SecRes}

Within the truncation scheme put forward in App. \ref{AppA} we can compute the contact as a function of the crossover parameters $\mu$, $T$, and $a$. Moreover, the high momentum factorization of the self-energy can be shown explicitly by solving the flow equation for $\Sigma_\psi(P)$ for different values of $P$. In order to translate the results expressed in terms of the chemical potential for the density, the equation of state $P(\mu,T)$ has to be applied. Since the density (and, iteratively, the contact itself) receives substantial contributions from the contact term in the fermion self-energy, fully self-consistent results can only be obtained from a self-consistent treatment of the Tan term in the flow equations. Here we restrict to an analysis of the qualitative behavior of the contact and do not aim at quantitative precision.

At zero temperature, the expression for the contact on the BEC side can be derived from the Lee--Huang--Yang (LHY) equation of state (\ref{C-2}) and the adiabatic sweep theorem (\ref{1-2}). We recapitulate this derivation in App. \ref{AppC}, where we also give the corresponding expression in terms of $\mu_{\rm mb}=\mu-\vare_{\rm B}/2$, with (negative) binding energy $\vare_{\rm B}$ of the molecules.

\begin{figure}[t!]
 \centering
 \includegraphics[bb=0 0 249 176,scale=0.98,keepaspectratio=true]{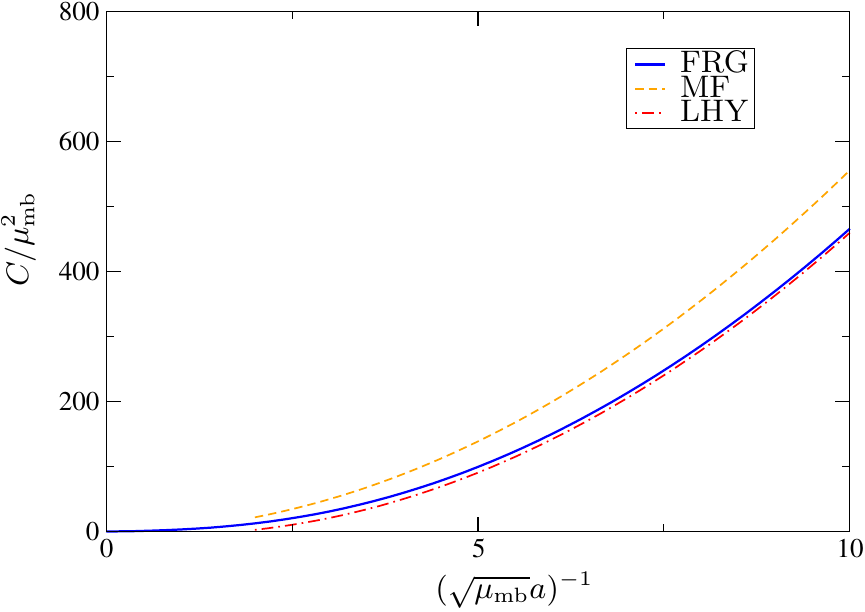}
 \caption{(Color online) Zero temperature contact on the BEC side of the crossover. The many-body chemical potential is defined as $\mu_{\rm mb}=\mu-\varepsilon_{\rm B}/2$ and thus a positive quantity. (See for instance Eq. (\ref{C-5}) in the context of a weakly interacting Bose gas.) The FRG treatment captures the  Lee--Huang--Yang (LHY) correction, see. Eq. (\ref{C-9}). Mean field theory (MF) is shown by the dashed curve.} 
 \label{FigBEC}
\end{figure}

\begin{figure}[t!]
 \centering
 \includegraphics[bb=0 0 240 176,scale=1.0,keepaspectratio=true]{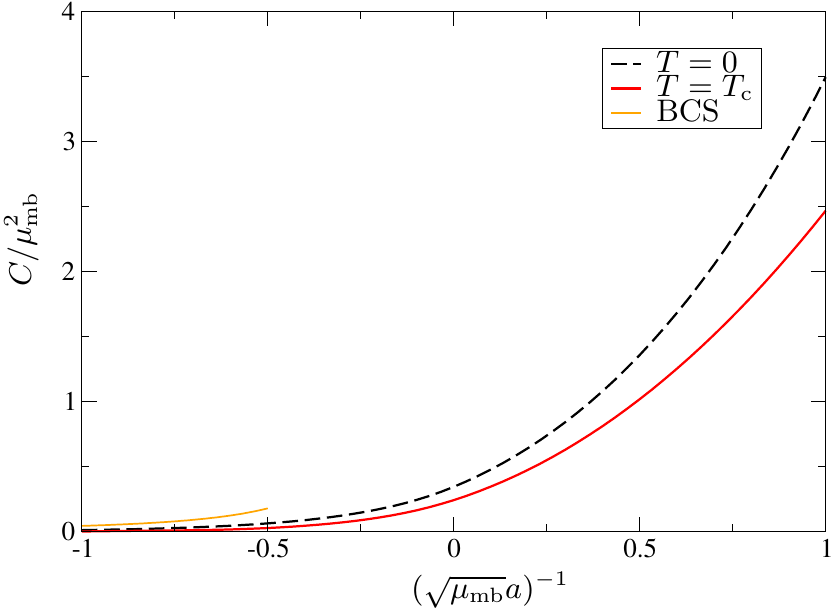}
 \caption{(Color online) The contact close to resonance as a function of the scattering length for both $T=0$ and $T=T_{\rm c}(a,\mu)$. The labels of the axes are analogous to Fig. \ref{FigBEC}. At unitarity we obtain $C(T=0)/\mu^2=0.34$ and $C(T=T_{\rm c})/\mu^2=0.24$. Far on the BCS side our present truncation becomes inappropriate as is discussed in App. \ref{AppAtoms}. For better visibility we show the asymptotic BCS value up to $(\sqrt{\mu}a)^{-1}=-0.5$, which is already beyond the applicability of BCS theory.}
 \label{FigUnitT}
\end{figure}

The result of the integration of the renormalization group equations at zero temperature on the BEC side is given in Fig. \ref{FigBEC}. We find excellent agreement with the prediction from LHY theory, whereas the mean field curve deviates substantially. From Eq. (\ref{B-21}) it is apparent that the LHY correction, which is reproduced in the equation of state on the BEC side as well, is also visible in the contact, because both share a common flow equation. The nontrivial renormalization of the prefactor in Eq. (\ref{B-21}) ensures the result to be beyond mean field.

\begin{figure}[h!]
 \centering
 \includegraphics[bb=0 0 249 176,scale=0.98,keepaspectratio=true]{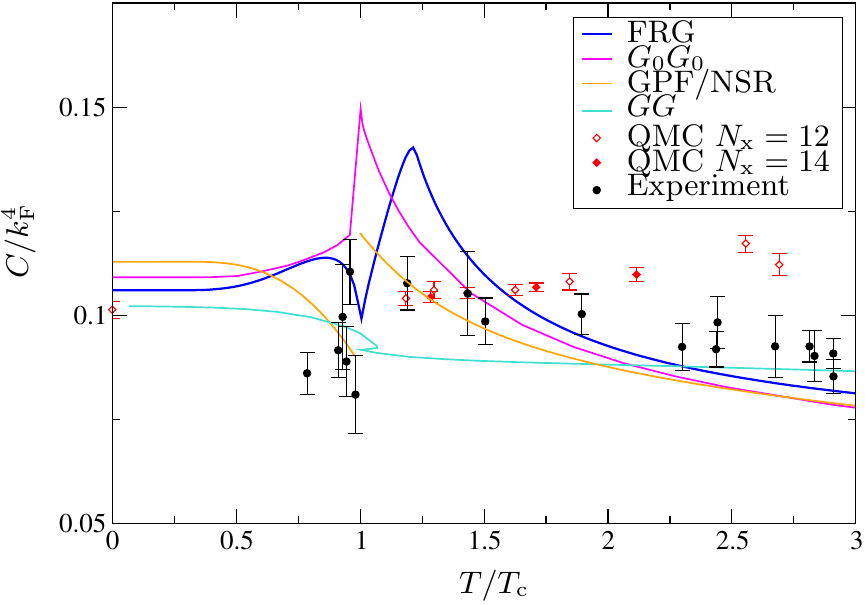}
 \caption{(Color online) The contact of the Unitary Fermi gas normalized by $k_{\rm F}^4$ with Fermi momentum $k_{\rm F}=(3\pi^2n)^{1/3}$. We compare predictions from Functional Renormalization Group (FRG, this work, $T_{\rm c}/T_{\rm F}=0.276$), non-self-consistent t-matrix theory ($G_0G_0$, \cite{PhysRevA.82.021605}, $T_{\rm c}/T_{\rm F}=0.242$), Gaussian pair fluctuations and Nozi\`{e}res--Schmitt-Rink theory (GPF/NSR, \cite{1367-2630-13-3-035007}, $T_{\rm c}/T_{\rm F}=0.235$), self-consistent t-matrix theory ($GG$, \cite{Enss2011770}, $T_{\rm c}/T_{\rm F}=0.15$) and Quantum Monte Carlo calculations (QMC, \cite{PhysRevLett.106.205302}, $T_{\rm c}/T_{\rm F}=0.15$) for lattice sizes $N_x=12, 14$. In this list, the brackets indicate the label in the plot, the corresponding reference and the chosen value for the critical temperature. For the experimental data of Ref. \cite{JinContact} we employed $T_{\rm c}/T_{\rm F}=0.16$, which suffices here to obtain a qualitative comparison of the data.}
 \label{FigCkF}
\end{figure}

As we approach unitarity from the BEC side, we leave the perturbative regime and the contact is no longer described by the LHY expression. The FRG result within the truncation of this work is shown in Fig. \ref{FigUnitT}. For $a^{-1}=0$ we find $C/\mu^2= 0.34$ and $C/k_{\rm F}^4=0.11$ at zero temperature. The Bertsch parameter within this approximation is $\xi=0.55$. We observe the contact to be a monotonous function of the crossover parameter $(\sqrt{\mu_{\rm mb}}a)^{-1}$, or, equivalently, $(k_{\rm F}a)^{-1}$.

For negative scattering lengths we find our zero temperature results to be far below the BCS prediction. The reason for the failure of the present truncation is that the momentum dependence of the boson propagator is not well-approximated by a derivative expansion on the BCS side, although momentum independent observables like the equation of state are described correctly. This becomes transparent in the derivation of the relation $C_{\rm BCS}=4 \pi^2n^2a^2$ from the Schwinger--Dyson equation in Eq. (\ref{sd26}), where we explicitly use the momentum dependence of the bosonic self-energy $\Sigma_\vphi(Q)$.   In App. \ref{AppAtoms} we discuss how the contact on the BCS side is more easily derived in a purely fermionic language.

\begin{figure}[t!]
 \centering
 \includegraphics[bb=0 0 245 174,scale=1.0,keepaspectratio=true]{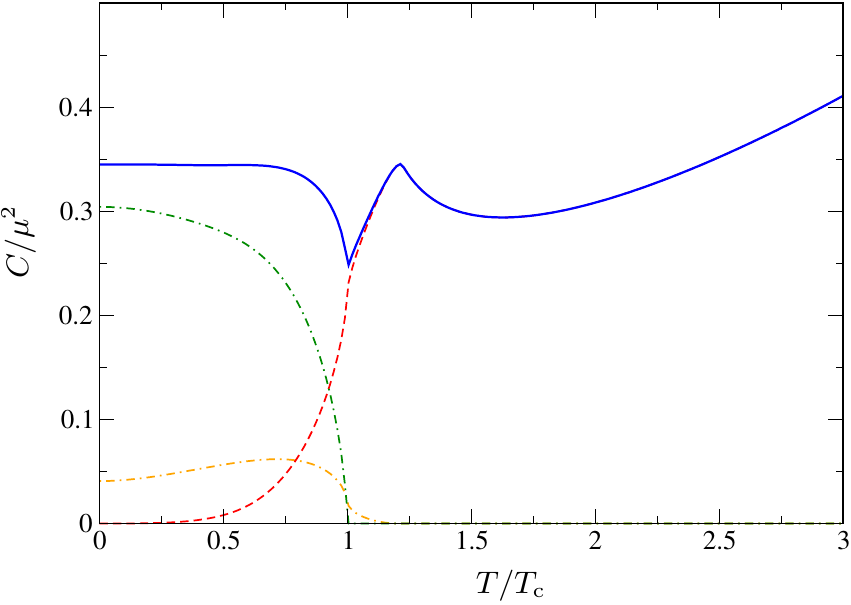}
 \caption{(Color online) The blue solid line shows the temperature dependence of the contact normalized by the chemical potential  for the Unitary Fermi gas. We observe a decrease of $C/\mu^2$ as we approach the critical temperature from below, resulting in a sharp dip at $T_{\rm c}$.  The function is monotonic for $T/T_{\rm c} \gtrsim 1.75$. The green, red, and orange curves correspond to different contributions to the contact and are explained in Eqs. (\ref{3-35})--(\ref{3-37}).}
 \label{FigCT}
\end{figure}

In Fig. \ref{FigUnitT} we also show the value of the critical contact $C(T_{\rm c},a)$ in the crossover. We find the corresponding value always to be below the zero temperature value. The full temperature dependence of the contact of the unitary Fermi gas is shown in Figs. \ref{FigCkF} and \ref{FigCT}. We observe a sharp dip at the critical temperature, hence $C(0) > C(T_{\rm c})$ in Fig. \ref{FigUnitT}. Since the contact is related to a first derivative of the energy (or pressure) according to the adiabatic sweep theorem (\ref{1-2}), it has to be continuous at $T_{\rm c}$ as a result of the second order nature of the phase transition. We confirm this behavior in our results with a critical contact parameter $C(T=T_{\rm c})/k_{\rm F}^4 =0.11$. The contact $C/k_{\rm F}^4$ shows a maximum above $T_{\rm c}$.

Fig. \ref{FigCkF} also compares our result for $C/k_{\rm F}^4$ to other theoretical approaches and to a recent experimental measurement of the homogeneous contact. Due to the disagreement of predictions for $T_{\rm c}/T_{\rm F}$ from different theoretical methods, we have rescaled the abscissa by the corresponding critical temperatures. This allows to compare the qualitative features of the temperature dependence like monotony or location of peaks and minima.  In order to relate the contact $C$ in Fig. \ref{FigCkF} to an extensive contact $\bar{C}=CV$ with volume $V$ we use $\bar{C}/N k_{\rm F} = 3 \pi^2 C/k_{\rm F}^4$, see our discussion of the normalization in the introduction. The Fermi momentum of the FRG data in Fig. \ref{FigCkF} is not corrected due to the high momentum contribution to the particle number density. Hence, $k_{\rm F}$ will in general be larger than plotted here.

We find largely different predictions for the temperature dependence of the contact in the critical region. This indicates a sensitivity of this observable with respect to approximations in theoretical calculations, which makes further investigation even more interesting. Note that for higher temperatures, the second and third order virial expansions of Ref. \cite{1367-2630-13-3-035007} allow for a solid comparison of the temperature dependence of the contact. However, we focus here on the region around $T_{\rm c}$, which is well-captured by our truncation of the effective action.

From Eq. (\ref{3-29}) we observe that the contact receives contributions from different terms in the flow equation. These are important in distinct regimes of the system. To visualize this, we split up the flow of $C_k$ into three parts according to $\partial_t C_k = \partial_t C_k^{(1)} + \partial_t C_k^{(2)} + \partial_t C_k^{(3)}$ with
\begin{align}
 \label{3-35} \partial_t C_k^{(1)} &= \frac{h^2_\phi}{4} (\eta_A \rho_{0,k}+\partial_t \rho_{0,k}),\\
  \label{3-36} \partial_t C_k^{(2)} &=\frac{2^{d/2+1}v_d }{d} \frac{h_\phi^2}{S^2}k^{d+2}\Bigl(1-\frac{\eta}{d+2}\Bigr)N_B'(\omega_{\phi,k}),\\
  \nonumber \partial_t C_k^{(3)} &=\frac{2^{d/2+1}v_d }{d} \frac{h_\phi^2}{S^2}k^{d+2}\Bigl(1-\frac{\eta}{d+2}\Bigr)\\
 &\label{3-37} \Bigl(-\frac{\lambda^2\rho_0^2}{S^2\omega_{\phi,k}^3}\Bigr)\Bigl(\frac{1}{2}+N_B(\omega_{\phi,k})-\omega_{\phi,k}N_B'(\omega_{\phi,k})\Bigr).
\end{align}
The only term which persists in the  stages of the flow where $\rho_{0,k}=0$, is $C^{(2)}_k$. Therefore, it is the leading contribution above $T_{\rm c}$ (red dashed line in Fig. \ref{FigCT}). Both $C^{(1)}_k$ and $C^{(3)}_k$ start to build up in the precondensation phase. However, $C^{(3)}_k$ is never really large (orange dashed-dotted line in Fig. \ref{FigCT}). For small $T$, the contribution from $C^{(1)}_k$ dominates (green dashed-dotted line in Fig. \ref{FigCT}). Above $T_{\rm c}$ this contribution is negligible. In the zero temperature limit, the term $C^{(1)}_k$ becomes most important. This can be understood easily from Eq. (\ref{B-16}), where we identify this term as the contribution from condensed bosons to the density.

\begin{figure}[t!]
 \centering
 \includegraphics[bb=0 0 254 173,scale=0.96,keepaspectratio=true]{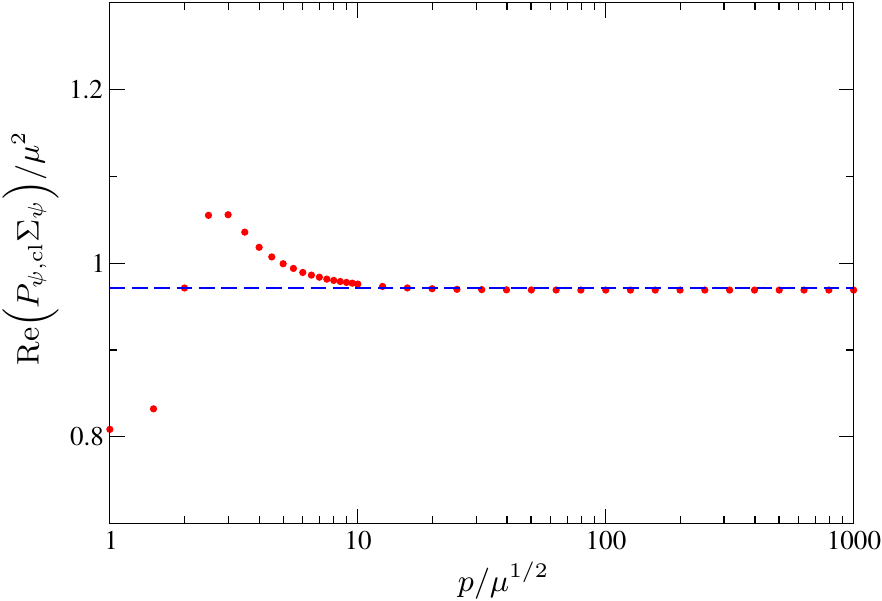}
 \caption{(Color online) The asymptotic approach of the contact for the unitary Fermi gas at the critical temperature. We plot the real part of $P_{\psi,\rm cl}(-P)\Sigma_\psi(P)$ with $P_{\psi,\rm cl}(-P) = -\rmi p_0 +\vec{p}^2-\mu$ and $p_0=\pi T$ (red dots). In accordance with formula (\ref{1-5}), the factorization at large momenta leads to the approach of the constant value $4C$ (blue dashed line). For very low momenta our perturbative treatment of the fermion propagator becomes quantitatively less accurate, but still the deviations do not exceed 20\% even for $p\to 0$. We restrict the self-energy here to the diagram in Fig. \ref{FigFermFlow} where the external momentum $P$ appears in the fermion line and which is responsible for the contact term. This effectively adds the constant $\delta \mu$ to $\Sigma_\psi$ for large momenta.}
 \label{FIGapproach}
\end{figure}

\begin{figure}[t!]
 \centering
 \includegraphics[bb=0 0 249 173,scale=0.98,keepaspectratio=true]{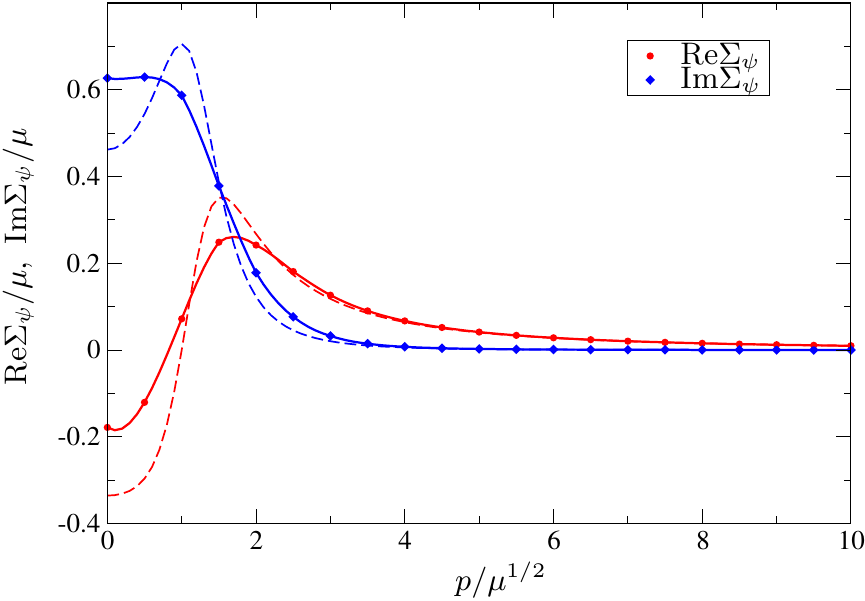}
 \caption{(Color online) The fermion self-energy $\Sigma_{\psi,k=0}(P)$ at $a^{-1}=0$ and $T=T_{\rm c}$ computed from the first diagram in Fig. \ref{FigFermFlow}. We evaluate the function for $p_0=\pi T$ and find the large momentum behavior to be a reasonable approximation for both the real and imaginary parts even at low momenta. The asymptotic form (\ref{1-5}) is shown by a dashed line. The self-energy vanishes for large momenta since we effectively added the constant $\delta \mu$ for $p\to \infty$ by neglecting the second diagram in Fig. \ref{FigFermFlow}.}
 \label{FIGFullSigmaTc}
\end{figure}

We already addressed the question whether the scaling formula (\ref{1-5}) can be applied for a large part of the momenta or only yields an asymptotic, but practically irrelevant contribution. For this purpose we solve the flow equation for the self-energy $\Sigma_{\psi,k}(P)$ on a grid of $P$-values according to Eq. (\ref{B-9}). Therein we restrict to the first integral, which corresponds to the diagram in Fig. \ref{FigFermFlow} where the external momentum appears in the fermion line. Only this diagram contributes to the contact.  The universal regime of validity is expected to be large for the Unitary Fermi gas. In Figs. \ref{FIGapproach} and \ref{FIGFullSigmaTc} we underline this statement at $T=T_{\rm c}$ and $a^{-1}=0$.

Since the self-energy $\Sigma_\psi(P)$ is a complex valued function of $P=(p_0,\vec{p})$, we gain information about the high momentum behavior from plotting both the real and imaginary part at the lowest possible fermionic Matsubara frequency $p_0=\pi T$ as a function of $|\vec{p}|$. To see the asymptotic approach of the form $P_{\psi,\rm cl}(-P)\Sigma_\psi \simeq 4C$ we plot the real part of this particular combination in Fig. \ref{FIGapproach}. The imaginary part of this product vanishes for $p\to \infty$, showing that the contact indeed is real-valued.

We plot  $\mbox{Re}\Sigma_\psi(P)$ and $\mbox{Im}\Sigma_\psi(P)$ in Fig. \ref{FIGFullSigmaTc} for the same set of parameters as before. We find that the scaling form is a good description for all momenta at the critical temperature. Although this does not come unexpected for a scale invariant, critical system, this behavior could be a relict of our perturbative treatment of the propagator for small momenta, where this is not necessarily a valid assumption. Further improvement of the truncation and iterative solution of the flow equation will shed light on the reliability of this finding, but is beyond the scope of the present work.

\section{Discussion \& Outlook}
\label{SecDis}

In this work we derive the contact in the BCS-BEC crossover from a universal factorization of the large momentum part of the fermion self-energy. The analysis is built on exact Schwinger--Dyson equations and an exact Functional Renormalization Group equation. Approximative solutions for both methods allow for a computation of the contact $C(\mu,T,a)$ for arbitrary values of the parameters. In the limiting perturbative cases the expected relations $C_{\rm BEC}=4\pi n/a$ and $C_{\rm BCS}=4 \pi^2 n^2 a^2$ can be obtained analytically.

The factorization (\ref{1-5}) of the self-energy into a universal part and the momentum independent many-body contact can be shown explicitly by solving the renormalization group flow equation for $\Sigma_{\psi,k}(P)$ for particular values of $P$. The regime of universal scaling turns out to be large for the Unitary Fermi gas. This has profound consequences for the density. Indeed, if we assume the scaling behavior of the self-energy to be valid for all momenta, we can estimate the contribution of high momentum particles to the total density in the symmetric phase by
\begin{align}
 \nonumber \delta n^{(C, \rm est)} = 8C \int \frac{\mbox{d}^3p}{(2\pi)^3} &\Bigl(\frac{1-2N_F(p^2-\mu)}{4(p^2-\mu)^2}\\
 \label{4-1} &+\frac{N_F'(p^2-\mu)}{2(p^2-\mu)}\Bigr)
\end{align}
see Eq. (\ref{sd14}) above. 

This density correction from high momentum particles  has a direct impact on the value of the critical temperature For the truncation employed in this paper and neglecting the Tan correction one finds for the critical temperature at unitarity $T_{\rm c}/T_{\rm F}=0.276$ and $T_{\rm c}/\mu=0.44$. The discrepancy to other methods is larger in a normalization involving the density, i.e. $T_{\rm c}/T_{\rm F}$. Part of the mismatch is likely to be found in the normalization with an incomplete density where substantial contributions are not included. Assuming formula (\ref{4-1}) to be valid, we find a correction of the density $\delta n^{(C, \rm est)}/(n^{(0)}+\delta n^{(C, \rm est)})=0.29$, resulting in a corrected value $T_{\rm c}/T_{\rm F}=0.22$. Here $n^{(0)}$ is the density which is obtained from the standard truncation without contact as it is described in App. \ref{AppA}.

The suppression of the ratio $T_{\rm c}/T_{\rm F}$ due to the high momentum tail of the particle density distribution also sheds light on the success in computing the critical temperature with methods which take into account the full momentum dependence of the propagators, e.g. Monte Carlo simulations or self-consistent gap equations. However, having singled out this important contribution to the fermion propagator, an effective model which takes into account the Tan term might give reasonable results with less numerical effort and, furthermore, help improving our understanding of the physics of the BCS-BEC crossover.

To further improve our truncation we can feed back the effect of the contact on the remaining running couplings. Indeed, so far we only have evaluated $\partial_t C_k$ on the solution of the truncation with classical fermion propagator. However, as a correction to the fermion propagator, the self-energy results in a modified flow of, e.g., the effective potential. A different running of the boson mass term $m^2_{\phi,k}$ results in a different phase diagram and hence a corrected value for $T_{\rm c}/\mu$. 

To close the present truncation of the FRG equations up to order $P_{\rm cl}^{-1}(Q)$-effects in the self-energies, we have to include also the bosonic contact, which arises in the high momentum limit of the bosonic self-energy. The relevance of the contact corrections is most pronounced for the pressure of the system, which is minus the effective potential and thus only includes a single propagator in the corresponding flow equation. However, we have seen on the example of the density that also corrections to first derivatives of the effective potential, i.e. density $n_k$, boson mass $m^2_{\phi,k}$ and condensate density $\rho_{0,k}$, can be substantial. Hence a closed set-up for improvement of our truncation needs to incorporate the fermionic and bosonic contact as a correction to the flow of these couplings. This can be achieved in a perturbative and iterative fashion.

Our findings suggest that a rather simple approximation to the full inverse fermion propagator
\begin{align}
 \nonumber P_\psi(Q) = &\mbox{ }\rmi Z_\psi q_0 +A_\psi q^2-\mu -\delta\mu\\
 &+\frac{4C(\rmi q_0+q^2-\mu+R_k(Q))}{q_0^2+(q^2-\mu+R_k(Q))^2+\kappa}
\end{align}
with $k$-dependent couplings $Z_\psi$, $A_\psi$, $\delta\mu$, $C$, and $\kappa$, and infrared cutoff $R_k(Q)$, combined with a suitable generalization of the inverse boson propagator, will lead to a substantial improvement of the quantitative precision in the FRG treatment of strongly interacting fermionic systems. We observe that in the superfluid  regime the inverse fermion propagator has an additional off-diagonal contribution $\sim h_\phi \sqrt{\rho_0}$. This regularizes the momentum integrals, such that the explicit regulator $\sim \kappa$ may not be needed. The occupation number corresponding to this ansatz reads
\begin{align}
 \label{dichte2} n_{\psi,\vec{q}\sigma} = - \Bigl( T \sum_n \frac{P_\psi(Q)}{P_\psi(Q)P_\psi(-Q)+h^2_\phi\rho_0} -\frac{1}{2}\Bigr)
\end{align}
with $k$-dependent density $n_k=2\int_{\vec{q}}n_{\psi,\vec{q}\sigma}$. The flowing density, and therefore the total density $n=n_{k=0}$, can thus be inferred from the flowing couplings $Z_\psi$, $A_\psi$, $\delta\mu$, $C$ and $\kappa$.

\begin{center}
 \textbf{Acknowledgements}
\end{center}

\noindent We thank T.~Enss, S.~Moroz, D.~Goncalves-Netto, M.~Scherer, D.~Schnoerr, and W.~Zwerger for 
discussions and helpful comments and P.~Drummond, J. E.~Drut, T.~Enss, D.~Jin, F.~Palestini, P.~Pieri, and Y.~Sagi
for providing us with their data. S.~D. acknowledges support by the Austrian Science Fund (FWF) through
SFB FoQuS (FWF Project No. F4006-N16) and the START grant Y 581-N16. I.~B.\ acknowledges funding
from the Graduate Academy Heidelberg. This work is supported by the
Helmholtz Alliance HA216/EMMI.

\begin{appendix}
\section{Contact in the perturbative regime}
\label{AppSDE}

To exemplify our statements in Sec. \ref{SecSDE}, we consider here the perturbative BEC and BCS regimes, where the integrals in Eqs. (\ref{sd5d}) and (\ref{sd5h}) can be performed analytically. For weak interactions, the largest physical scale is given by $k_{\rm ph}=a^{-1}$. To study the contact in the whole crossover we employ the Functional Renormalization Group in Sec. \ref{SecFRG}. One of the merits of this method is that all expressions are automatically renormalized. 

On the BEC and BCS sides of the crossover, the fermion propagator only gets weakly dressed because either the fermion or the boson propagator is gapped. This leads to a suppression of the loop-integral originating from the diagram shown in Fig. \ref{FigFermSDE}. Hence we can always treat the fermion self-energy perturbatively in these regimes. We identify the large momentum shift of the chemical potential as
\begin{align}
 \nonumber - \delta \mu &= h_\vphi^2 \int_{\vec{p}}^\Lambda T \sum_n \frac{1}{(\rmi(p_0-q_0)+p^2-\mu)(\nu +\rmi \vare p_0)}\\
 \label{sd16} &= h_\vphi^2\int_{\vec{p}} \frac{-N_F(p^2-\mu)-N_B(\nu/\vare)}{\vare(\rmi q_0-p^2+\mu)+\nu} =-\frac{h_\vphi^2}{\nu}n_{\psi,\sigma}
\end{align}
on the BCS side. The Bose function is denoted by $N_B(z)=(e^{z/T}-1)^{-1}$. In the last expression we take the limit $\vare \to 0^+$. The $q_0$-dependence of $\delta \mu$ defined in Eq. (\ref{sd5h}) is seen to vanish.

\begin{figure}[t!]
 \centering
 \includegraphics[scale=0.9,keepaspectratio=true]{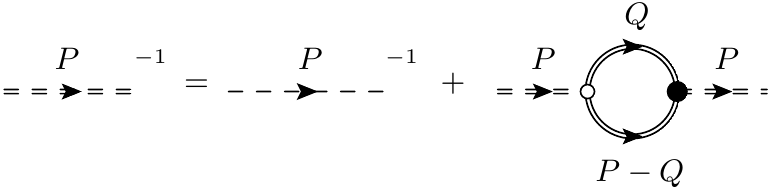}
 \caption{Schwinger--Dyson equation for the inverse boson propagator. The notation is chosen as in Fig. \ref{FigFermSDE}.}
 \label{FigBosSDE}
\end{figure}

We first consider the BEC limit, where $\mu \simeq -1/a^2$ is large and negative. The shift of the chemical potential vanishes due to the suppression of the Fermi function $N_F(p^2-\mu)$ in the integral in Eq. (\ref{sd16}). The inverse boson propagator is derived from the Schwinger--Dyson equation  \cite{Diehl:2005ae} for the bosonic self-energy
\begin{align}
 \nonumber \Sigma_\vphi(Q) &= P_{\vphi}(Q) - P_{\vphi, \rm cl}(Q)\\
 \label{sd20} &= \delta\nu_\Lambda -h_\vphi^2\int_P\frac{1}{P_\psi(Q-P)P_\psi(P)},
\end{align}
shown diagrammatically in Fig. \ref{FigBosSDE}. The counterterm is given by
\begin{align}
 \label{sd21} \delta \nu_\Lambda = \frac{h_\vphi^2\Lambda}{4\pi^2}.
\end{align}
In the perturbative regime we can replace the full fermion propagators in the loop-integral by the classical ones. The Matsubara summation and angular integration can then be evaluated analytically. We do not need the full expression but only note that
\begin{align}
  \label{sd22}P_{\vphi,\rm BEC} (Q) = \nu +\frac{h_\vphi^2}{8\pi} \sqrt{\frac{\rmi q_0}{2}+\frac{q^2}{4}-\mu}\simeq Z_\vphi \Bigl(\rmi q_0+\frac{q^2}{2}\Bigr)
\end{align}
for $ |q_0|,q^2\ll a^{-2}$ in the BEC regime. (We used $-h_\vphi^2/\nu=8\pi a$ and $\mu = -1/a^2$) The boson propagator resembles particles with classical dispersion relation $\omega_q=q^2/2M_\vphi$ and mass $M_\vphi = 2M=1$. The wave function renormalization constant is $Z_\vphi=h_\vphi^2a/32\pi$.

Inserting the boson propagator from Eq. (\ref{sd22}) into Eq. (\ref{sd5e}) we arrive at
\begin{align}
 \nonumber &C_{\rm BEC} = \frac{8 \pi}{a} \int_{p^2<k_{\rm tr}^2} T \sum_{n} \Bigl(\frac{1}{\rmi p_0+p^2/2}-\frac{1}{\rmi p_0+k^2_{\rm tr}}\Bigr)\\
 \label{sd23} &= \frac{8\pi}{a} \int_{p^2<k_{\rm tr}^2} \Bigl(N_B(p^2/2) - N_B(k_{\rm tr}^2) \Bigr) = \frac{8 \pi n_{\vphi, \rm cl}}{a} ,
\end{align}
where we used $k_{\rm ph} \gg T^{1/2}$ and $n_{\vphi,\rm cl}$ defines the number density of boson with classical dispersion relation from Eq. (\ref{sd5f}). Due to the equation of state $n_{\vphi, \rm cl}=n/2$ on the BEC side of the crossover we conclude
\begin{align}
 \label{sd24} C_{\rm BEC} = \frac{4 \pi n}{a}
\end{align}
as expected. The perturbative expression for the contact starting from the equation of state at zero temperature is recapitulated in App. \ref{AppC}.

In the BCS regime the bosons are resonant excitations and $n_{\vphi,\rm cl}=0$. The corresponding formula for the contact is most easily derived from inserting the Schwinger--Dyson Eq. (\ref{sd20}) into formula (\ref{sd5e}). Therein, the boson self-energy $\Sigma_\vphi \propto h_\vphi^2$ can be treated perturbatively, since it is small in comparison to the boson gap $\nu$ due to the small scattering length $a=-h_\vphi^2/8 \pi\nu$. We then find
\begin{align}
 \nonumber C_{\rm BCS} &= -\frac{h^2_\vphi}{4} T \sum_n \int_{p^2 < k_{\rm tr}^2} \frac{\Sigma_\vphi(P)-\Sigma_\vphi(p_0,k_{\rm tr}^2)}{(\nu+\rmi \vare p_0)^2}\\
 \nonumber &= \frac{h^4_\vphi}{4\nu} T \sum_n \int_{p^2 < k_{\rm tr}^2} \int_K \frac{1}{P_{\psi,\rm cl}(K)(\nu+\rmi \vare p_0)}\\
 \nonumber &\times\biggl(\frac{1}{P_{\psi,\rm cl}(P-K)}-\frac{1}{P_{\psi,\rm cl}(p_0-k_0,(k_{\rm tr}-k)^2)}\biggr)\\
 \nonumber& = - \frac{h^4_\vphi}{4\nu} \int_K \int_{p^2<k_{\rm tr}^2} \frac{N_F((\vec{p}-\vec{k})^2-\mu)}{P_{\psi, \rm cl}(K)(\nu + \rmi \vare k_0)}\\
 \nonumber &= -\frac{h^4_\vphi}{4\nu} n_{\psi,\sigma}\int_K \frac{1}{P_{\psi,\rm cl}(K)(\nu+\rmi \vare k_0)} \\
 \label{sd25} &= -\frac{h^4_\vphi}{4\nu} n_{\psi,\sigma}\int_{\vec{k}} \frac{N_F(k^2-\mu)-N_F(\nu/\vare)}{\vare(k^2-\mu)-\nu} = \frac{h^4_\vphi}{4\nu^2}n_{\psi,\sigma}^2.
\end{align}
We again applied the limit $\vare \to 0^+$. With the BCS equation of state $n_{\psi,\sigma}=n/2$ we arrive at
\begin{align}
 \label{sd26} C_{\rm BCS} = 4 \pi^2 a^2n^2.
\end{align}
This agrees with Eq. (\ref{C-13}).

\section{Contact in the atomic phase}
\label{AppAtoms}

Deep in the atomic phase, i.e. for high temperatures or on the BCS side of the crossover, bosons are excited
resonantly. For this purpose, an ansatz $P_\phi(Q) \approx P_{\phi,\rm cl}(Q)$ fails to capture the right
behavior of the contact. This is, however, not a failure of the Yukawa model, but rather shows the need for
keeping higher order terms in the bosonic self-energy. We have seen in Eq. (\ref{sd26}) that the  BCS result $C_{\rm BCS}=4 \pi^2a^2n^2$ can
be obtained from the full boson propagator. Our truncation for solving the FRG equation for the fermion propagator, as it is described in Sec. \ref{SecFRG} and App. \ref{AppA}, cannot resolve this behavior, since the bosonic self-energy is only incorporated by the wave function renormalization coefficients $A_\phi$, $Z_\phi$ and the boson mass term $m^2_\phi$. This is made explicit in Eq. (\ref{A-3b}).

\begin{figure}[t!]
 \centering
 \includegraphics[bb=0 0 286 45,scale=0.8,keepaspectratio=true]{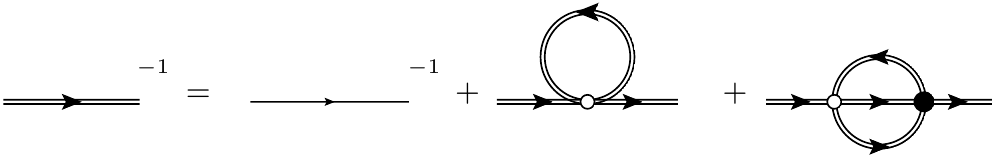}
 \caption{Schwinger--Dyson equation for the inverse fermion propagator in a purely fermionic theory. Again, a single or double line denotes a classical or full propagator. The empty (filled) dot corresponds to a classical (full) four-fermion vertex. We can combine the last two terms such that the correction is only given by the tadpole diagram but with a momentum dependent effective vertex.}
 \label{FigVierFermSDE}
\end{figure}

The contact in the atomic phase can be obtained within a purely fermionic language.
Then the momentum dependence of the related Schwinger--Dyson
equation for the propagator is solely encoded in the two-loop diagram
shown in Fig. \ref{FigVierFermSDE}, because the tadpole diagram contains the classical
vertex, which is momentum independent. Equivalently, we can express the correction in terms
of a tadpole diagram with full four-fermion vertex. A (2PI-)resummed form of the latter is given by  
\begin{equation}
 \label{X-1} \lambda_{\psi}(P) = \frac{\lambda_\psi}{1-\lambda_\psi\Bigl(\Pi(P)-\Pi_\Lambda(P)\Bigr)} 
\end{equation}
with $\Pi(Q)$ being defined in Eq. (\ref{sd25}) and $\Pi_\Lambda(P)$ chosen as a suitable subtraction at large momentum. Neglecting renormalization issues, we find for large external momenta $p^2$ to second order in $\lambda_\psi$ that
\begin{align}
 \nonumber \Sigma_\psi(P) &= -\int_Q \frac{\lambda_\psi(Q)}{P_\psi(Q-P)} \\
 \label{X-2} &\simeq \frac{1}{P_{\psi,\rm cl}(-P)} \int_Q \lambda_\psi^2 \Bigl(\Pi(Q)-\Pi_\Lambda(Q)\Bigr)
\end{align}
Employing $\lambda_\psi =- 8\pi a$ we recover the integral in Eq. (\ref{sd25}) and thus $C_{\rm BCS}=4 \pi^2a^2 n^2$. The FRG allows to go beyond the bubble resummation in Eq. (\ref{X-2}), see e.g. Ref. \cite{Fukushima:2012xw}.

\section{Truncation of the effective average action}
\label{AppA}
In order to solve the flow equation (\ref{3-3}), we have to employ a truncation of the effective average action $\Gamma_k[\bar{\phi},\psi]$ which is on the one hand tractable from a computational point of view, but also contains the relevant physics of the underlying system. For this purpose we make the ansatz
\begin{align}
 \nonumber \Gamma_k[\bar{\phi},\psi] = \int_X \biggl( &\sum_{\sigma=1,2}\psi^*_\sigma(\partial_\tau-\nabla^2-\mu)\psi_\sigma \\
 \nonumber &+\bar{\phi}^*\Bigl(Z_k\partial_\tau-A_k\frac{\nabla^2}{2}\Bigr) \bar{\phi}+\bar{U}_k(\bar{\rho}) \\
 \label{A-1} &-\bar{h}_\phi(\bar{\phi}^*\psi_1\psi_2-\bar{\phi}\psi_1^*\psi_2^*)\biggr).
\end{align}
For a detailed discussion of this nonperturbative derivative expansion scheme see \cite{PhysRevA.76.021602,PhysRevA.76.053627,ANDP:ANDP201010458}. The Yukawa model applied to many-fermion systems has also been studied in Ref. \cite{Birse:2004ha}.  The effective average potential $\bar{U}_k(\bar{\rho},\mu)$ with $\bar{\rho}=\bar{\phi}^*\bar{\phi}$ is expanded in a power series around the physical values $\rho_{0,k}$ and $\mu$ according to
\begin{align}
\nonumber  &\bar{U}_k(\bar{\rho},\hat{\mu}) \\
 \nonumber &= \bar{m}^2_{\phi,k,\hat{\mu}}(\bar{\rho}-\bar{\rho}_{0,k})+\frac{\bar{\lambda}_{\phi,k,\hat{\mu}}}{2}(\bar{\rho}-\bar{\rho}_{0,k})^2+\dots\\
 \nonumber &= \Bigl( \bar{m}^2_{\phi,k}(\bar{\rho}-\bar{\rho}_{0,k})+\bar{\alpha}_k(\hat{\mu}-\mu)(\bar{\rho}-\bar{\rho}_{0,k})+\dots\Bigr)\\
 \label{A-2} &+\Bigl(\frac{\bar{\lambda}_{\phi,k}}{2}(\bar{\rho}-\bar{\rho}_{0,k})^2+\dots\Bigr)+\dots
\end{align}
We neglect higher terms in this series and only keep track of the running of $\bar{m}^2_{\phi,k}$, $\bar{\lambda}_{\phi,k}$ and $\bar{\alpha}_k$. Going beyond this basic set of parameters yields higher quantitative precision. Note that we neglect a possible scale dependence of the Yukawa coupling $\bar{h}_\phi$. The initial conditions for the running couplings are set by $\Gamma_\Lambda=S$ to be
\begin{align}
 \nonumber &Z_\Lambda=A_\Lambda=1,\ \bar{m}^2_{\phi,\Lambda}=\nu_\Lambda-2\mu,\\
 \label{A-3} &\bar{\lambda}_{\phi,\Lambda}= 0,\ \bar{\alpha}_\Lambda=-2,\ \bar{\rho}_{0,\Lambda}=0.
\end{align}
The vacuum flow of $m^2_{\phi,k}$ allows for extracting the scattering length $a$ which is related to the detuning from resonance $\nu(B)$. Note that we define the detuning slightly different here than in Eq. (\ref{2-5}) since it does not contain the chemical potential.

The ansatz in Eq. (\ref{A-1}) takes into account the important low momentum modes of the boson propagator. To see this more clearly we expand the bosonic self-energy in the exact relation for the boson propagator
\begin{align}
 \nonumber &\bar{P}_\phi(Q) = \bar{P}_{\phi,\Lambda}(Q) + \bar{\Sigma}_\phi(q_0,q^2)\\
 \nonumber &= \bar{P}_{\phi,\Lambda}(Q) + \bar{\Sigma}(0,0) + \rmi q_0 \frac{\partial \bar{\Sigma}_\phi}{\partial \rmi q_0}(0,0)+q^2 \frac{\partial\bar{\Sigma}_\phi}{\partial q^2}(0,0)+\dots\\
 \label{A-3b} &= \rmi Z q_0 + \frac{1}{2} A q^2 + \bar{m}^2_\phi +\dots
\end{align}
with the corresponding identification of $Z$, $A$, and $\bar{m}^2_\phi$. We choose $Z_\Lambda$ and $A_\Lambda$ to be nonzero at the beginning of the flow. In fact, they are attracted immediately towards a fixed point with $Z_k=A_k=1$ in the early stages of the renormalization group low \cite{ANDP:ANDP201010458}. Thus, choosing $Z \neq 0$ immediately provides for the correct ultraviolet regularization of Eq. (\ref{sd2}). Of course, we can also choose $Z_\Lambda =\vare$ with sufficiently small $\vare >0$.

\begin{figure}[t!]
 \centering
 \includegraphics[bb=0 0 232 72,scale=1.0,keepaspectratio=true]{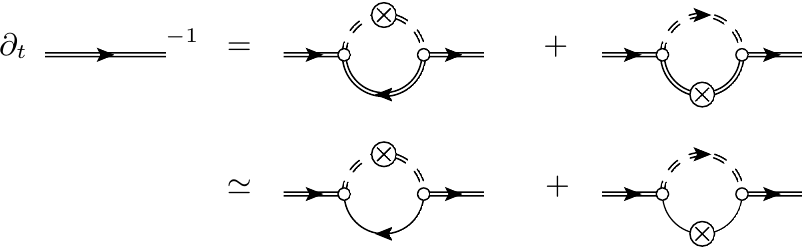}
 \caption{Flow equation for the fermion self-energy. The notation is chosen as in Fig. \ref{FigFermSDE}. The crossed circle indicates an insertion of $\dot{R}_k$ in the loop integral.}
 \label{FigFermFlow}
\end{figure}

The fermions are treated perturbatively but in a \emph{momentum resolved fashion}. Perturbatively here means that we neglect the feedback of the self-energy on the other running couplings. Of course, they can be implemented iteratively, thus enhancing the quantitative precision of the results. We have an additional flow equation for the fermion self-energy shown in Fig. \ref{FigFermFlow}. In the second line of the figure we replaced the full fermion propagator by the microscopic one, in accordance with our perturbative treatment. Note that all quantities on the right hand side of the flow equation in Fig. \ref{FigFermFlow} are known to us and the fermionic self-energy can be readily integrated.

From Eq. (\ref{A-1}) we can read off the inverse propagators for the bosons and fermions at scale $k$, which are, respectively, given by
\begin{align}
 \label{A-6} \bar{P}_{\phi,k}(Q) &= \rmi Z_k q_0 + A_k q^2/2+\bar{m}^2_{\phi,k},\\
 \label{A-7} P_{\psi,\Lambda}(Q) &=\rmi q_0 + q^2-\mu
\end{align}
in the symmetric phase. We employ regulator functions
\begin{align}
 \label{A-8} R_{\phi,k}(Q) &= A_k (k^2-q^2/2)\theta(k^2-q^2/2),\\
 \label{A-92} R_{\psi,k}(Q) &=(k^2\sgn\Bigl(\frac{q^2-\mu}{k^2}\Bigr)-(q^2-\mu))\theta(k^2-|q^2-\mu|),
\end{align}
which are optimized for a derivative expansion. Note that the functions $R_k(Q)$ do not depend on the frequency $q_0$, which facilitates a semi-analytical treatment of several of the loop-expressions encountered in the following. In addition, we introduce the abbreviations
\begin{align}
 \label{A-10} p_{\phi,k}(\vec{q}) &= \frac{1}{A} \bar{p}_{\phi,k}(\vec{q}) = q^2/2+m^2_{\phi,k} + R_{\phi,k}(q^2),\\
 \label{A-11} p_{\psi,k}(\vec{q}) &= q^2-\mu + R_{\psi,k}(q^2).
\end{align}
We normalize the running couplings by $A_k$ and define $S_k=Z_k/A_k$, $m^2_{\phi,k}=\bar{m}^2_{\phi,k}/A_k$, $\lambda_{\phi,k}=\bar{\lambda}_{\phi,k}/A_k^2$, $\rho_{0,k}=\bar{\rho}_{0,k}A_k$, and $h^2_k=\bar{h}^2_\phi/A_k$. Note that this introduces a $k$-dependence of the Yukawa coupling $h^2_k$. Moreover, we define the anomalous dimension $\eta_k=-\partial_t\ln A_k$, which substitutes the flow of $A_k$.

\section{Flow equation for the fermion self-energy and the contact}
\label{AppB}
The effective average action is parametrized in terms of the mean fields $\bar{\phi}_X$ and $\psi_X$, whereas the Schwinger functional is parametrized in terms of source fields. Both descriptions are equivalent and any mean field configuration, in particular with nonvanishing fermion expectation value, can be enforced by choosing an appropriate external source. Taking functional derivatives of $\Gamma_k[\bar{\phi},\psi]$ with respect to the fields $\bar{\phi}_X$ and $\psi_X$ serves for generating all one-particle irreducible correlation functions of the theory. In the same way, the flow equation (\ref{3-3}) serves for generating the flow of all correlation functions.

We consider the inverse fermion propagator for a constant (and real) bosonic background field $\bar{\rho}=\bar{\phi}^*\bar{\phi}$ and vanishing expectation value of the fermion mean field,
\begin{align}
 \label{B-1} G^{-1}_{\psi,k}(X,Y,\bar{\rho})= \frac{\stackrel{\rightarrow}{\delta}}{\delta \psi_{A}(X)} \Gamma_k \frac{\stackrel{\leftarrow}{\delta}}{\delta \psi_B(Y)}[\bar{\rho},0],
\end{align}
where $\psi_A$ and $\psi_B$ is any of the four indices $(\psi_1,\psi_2,\psi_1^*,\psi_2^*)$, such that we arrive at a $4\times4$--matrix. In a homogeneous situation, via Fourier transformation
\begin{align}
 \label{B-2} \int_{X,Y} e^{\rmi PX}e^{\rmi P'Y}G^{-1}_{\psi,k}(X,Y,\bar{\rho}) = \delta(P+P') G^{-1}_{\psi,k}(P,\bar{\rho})
\end{align}
we obtain the full inverse propagator $G^{-1}_{\psi,k}(P,\bar{\rho})$ in the presence of a background field $\bar{\rho}$.

The microscopic (or classical) inverse fermion propagator in the presence of a background field is given by
\begin{align}
 \label{B-3} G^{-1}_{\psi,\Lambda}(P,\bar{\rho}) = \begin{pmatrix} - \bar{h}_\phi \bar{\phi} \vare & -P_{\psi,\Lambda}(-P) \\ P_{\psi,\Lambda}(P) & \bar{h}_\phi \bar{\phi} \vare \end{pmatrix}.
\end{align}
Herein, the fully antisymmetric tensor with two indices is denoted by $\vare$. The physical self-energy is defined as the difference between the full (i.e., $k=0$) inverse propagator and the classical one:
\begin{align}
 \label{B-4} \Sigma(P) = G^{-1}_{\psi,k=0}(P) - G^{-1}_{\psi,\Lambda}(P).
\end{align}
Since for $k=0$ the bosonic field can have a nonvanishing expectation value $\rho_{0,k=0}$, we make this definition more explicit by writing
\begin{align}
 \label{B-5} \Sigma(P) = G^{-1}_{\psi,k=0}(P,\rho_{0,k=0}) - G^{-1}_{\psi,\Lambda}(P,\rho_{0,k=0}).
\end{align}
The most general form of the self-energy in the spin balanced case of equal chemical potentials for the hyperfine components can be parametrized by two complex-valued functions via
\begin{align}
\label{B-6} \Sigma(P) = \begin{pmatrix} \Sigma_{\psi,\rm an}(P)\vare & - \Sigma_\psi(-P) \\ \Sigma_\psi(P) & -(\Sigma_{\psi,\rm an}(P))^*\vare\end{pmatrix}.
\end{align}

Eq. (\ref{B-5}) is readily extended to a scale-dependent fermion self-energy by defining
\begin{align}
\label{B-7} \Sigma_{k}(P) =G^{-1}_{\psi,k}(P,\rho_{0,k}) - G^{-1}_{\psi,\Lambda}(P,\rho_{0,k}).
\end{align}
Evaluating the self-energy for each $k$ on the expectation value $\rho_{0,k}$ of the bosonic field properly takes into account the fluctuations on different scales. As a result, the flow equation of the $(\psi_1^*,\psi_1)$--component of $\Sigma_k$ (i.e. the normal contribution) is given by
\begin{align}
\nonumber  \partial_t \Sigma_{\psi,k}(P) =\mbox{ }&(\eta_A \rho_{0,k}+\partial_t\rho_{0,k})  \frac{\partial \Sigma_{k,\psi}}{\partial \rho}(P,\rho_{0,k})\\
 \label{B-8}&+\Bigl(\partial_t|_{\bar{\rho}}\Sigma_{\psi,k}\Bigr)(P,\rho_{0,k})
\end{align}
with RG-time $t=\ln(k/\Lambda)$ and normalized field  $\rho=\bar{\rho}A_k$. The derivative is performed for fixed $\bar{\rho}$. On the right hand side, the self-energy appears as a function of the background field, whereas the left hand side only depends on $k$ and $P$.

Within our truncation we have
\begin{widetext}
\begin{align}
 \nonumber \Bigl(\partial_t|_{\bar{\rho}}\Sigma_{\psi,k}\Bigr)(P,\rho_{0}) =\mbox{ }&h_\phi^2 \int_	Q \frac{\dot{R}_\phi(\vec{q}^2)}{A} \frac{\Bigl(S^2q_{0,B}^2-2(\rmi S q_{0,B}+\lambda\rho_0)(\lambda\rho_0+p_\phi(\vec{q}))-p_\phi^2(\vec{q})\Bigr)\Bigl(\rmi(p_0+q_{0,B})+p_\psi(\vec{q}+\vec{p})\Bigr)}{\mbox{det}^2_B(Q)\mbox{det}_F(Q+P)}\\
 \label{B-9} &-h_\phi^2 \int_Q \dot{R}_\psi(\vec{q}^2) \frac{\Bigl(\rmi S(p_0+q_{0,F})-\lambda\rho_0-p_\phi(\vec{q}+\vec{p})\Bigr)\Bigl(q_{0,F}^2+2\rmi q_{0,F}p_\psi(\vec{q})+h^2\rho_0-p_\psi^2(\vec{q})\Bigr)}{\mbox{det}_B(Q+P)\mbox{det}_F^2(Q)}
\end{align}
\end{widetext}
with $q_{0,B}=2\pi n T$ and $q_{0,F}=2\pi(n+1/2)T$ being bosonic or fermionic Matsubara frequencies, respectively, and $p_0=2\pi(m+1/2)T$. We introduced
\begin{align}
 \label{B-10} \mbox{det}_B(Q) &= S^2 q_{0,B}^2 + p_\phi(\vec{q})(p_\phi(\vec{q})+2\lambda\rho_0),\\
 \label{B-11} \mbox{det}_F(Q) &= q_{0,F}^2 + p_\psi^2(\vec{q}) +h_\phi^2\rho_0
\end{align}
and adopted the notation of Eqs. (\ref{A-10}) and (\ref{A-11}) for $p_{\phi,k}$ and $p_{\psi,k}$. The Matsubara summations can be performed analytically and we arrive at an explicit expression for the third term in the flow equation (\ref{B-8}).

The flow equation for $\Sigma_{\psi,k}(P)$ is valid for arbitrary values of $P$. For large $P$, the equation simplifies considerably and, eventually, allows to derive the renormalization group flow of the contact $C_k$. We restrict the following discussion to the first integral in Eq. (\ref{B-9}), which is responsible for the high momentum behavior. We have
\begin{align}
\label{B-13} \frac{\rmi(p_0+q_0)+p_{\psi}(Q+P)}{\mbox{det}_F(Q+P)} \simeq\frac{1}{-\rmi p_0+p^2-\mu}
\end{align}
for large $p^2$ and find 
\begin{align}
 \Bigl(\partial_t|_{\bar{\rho}}\Sigma_{\psi,k}\Bigr)(P,\bar{\rho}) \simeq \frac{(\partial_t|_{\bar{\rho}}\bar{c}_k)(\bar{\rho})}{P_{\psi,\Lambda}(-P)}
\end{align}
with
\begin{align}
 \nonumber &(\partial_t|_{\bar{\rho}}\bar{c}_k)(\bar{\rho})= h_\phi^2 \int_Q \frac{\dot{R}_\phi(\vec{q}^2)}{A}\\
 \nonumber & \times \frac{S^2q_{0,B}^2-2(\rmi S q_{0,B}+\lambda\rho_0)(\lambda\rho_0+p_\phi(\vec{q}))-p_\phi^2(\vec{q})}{\mbox{det}^2_B(Q)}\\
 \label{B-14} &=h_\phi^2 \int_{\vec{q}}\frac{\dot{R}_\phi}{A} T \sum_n \Bigl(\frac{1}{\mbox{det}_B(Q)}-\frac{2S^2\omega_{\phi,k}^2+2\lambda^2\rho_0^2}{\mbox{det}^2_B(Q)}\Bigr).
\end{align}
For the definition of $\omega_{\phi,k}$ see Eq. (\ref{3-30}).

We define the flowing contact according to
\begin{align}
 C_k = \frac{1}{4} c_k(\rho_{0,k}),
\end{align}
where $c_k(\rho):=\bar{c}_k(\bar{\rho})$ is expressed in terms of the normalized field $\rho=\bar{\rho} A_k$. Since
\begin{align}
 c(k,\rho) = \bar{c}(k,\bar{\rho}(\rho,k))
\end{align}
we have
\begin{align}
 \frac{\partial c}{\partial k} &= \frac{\partial \bar{c}}{\partial k} + \frac{\partial \bar{c}}{\partial \bar{\rho}} \cdot \left(\frac{\partial \bar{\rho}}{\partial k}\right)_\rho = \frac{\partial \bar{c}}{\partial k}+\frac{1}{k} \eta_A \bar{\rho} \frac{\partial \bar{c}}{\partial \bar{\rho}},\\
 \frac{\partial c}{\partial \rho} &= \frac{1}{A_k} \frac{\partial \bar{c}}{\partial \bar{\rho}}.
\end{align}
From these two relations we deduce the flow equation for $c_k(\rho)$ in the presence of the $k$--dependent background field $\rho=A_k\bar{\rho}$ to be given by
\begin{align}
 (\partial_t|_{\rho}c_k)(\bar{\rho})=(\partial_t|_{\bar{\rho}}\bar{c}_k)(\bar{\rho}=\bar{\rho}(\rho)) + \eta_A \rho \frac{\partial c_k}{\partial \rho}(\rho).
\end{align}
Thus we arrive at
\begin{align}
 \partial_t C_k =  \frac{1}{4}\Bigl( (\partial_t|_{\bar{\rho}}\bar{c}_k)(\rho_{0,k}) + (\eta_A \rho +\partial_t \rho_{0,k})\frac{\partial c_k}{\partial \rho}(\rho_{0,k})\Bigr).
\end{align}

We show that $\partial c/\partial \rho=h^2_{\phi}+O(\Sigma_\psi)$. For this purpose we consider the limit of classical fermion propagators $P_\psi(Q)=P_{\psi,\Lambda}(Q)=\rmi q_0+q^2-\mu$. For the momentum distribution per species we have
\begin{align}
 \nonumber n_{\psi,\vec{q}\sigma} &= - \Bigl( T\sum_n \frac{P_{\psi,\Lambda}(-Q)}{P_{\psi,\Lambda}(Q)P_{\psi,\Lambda}(-Q)+h^2 \rho}-\frac{1}{P_{\psi,\Lambda}(Q)}\Bigr)\\
 \nonumber &=T \sum_n \frac{h_\phi^2\rho}{P_{\psi,\Lambda}(Q)\Bigl[P_{\psi,\Lambda}(Q)P_{\psi,\Lambda}(-Q)+h_\phi^2\rho\Bigr]}\\
 &\simeq \frac{h_\phi^2\rho}{2\sqrt{q^4+h_\phi^2\rho}\Bigl(\sqrt{q^4+h_\phi^2\rho}+q^2\Bigr)} \simeq \frac{h_\phi^2 \rho}{4q^4}
\end{align}
for large $q$, see Eqs. (\ref{sd9}) and (\ref{dichte2}). Whereas the Fermi--Dirac distribution decays exponentially for large $q$, a $q^4$-tail arises from the presence of a bosonic background field with contact parameter $C=h_\phi^2 \rho/4$. This completes the proof of
\begin{align}
 \frac{\partial c}{\partial \rho}(\rho_{0,k}) = h^2_\phi +\mathcal{O}(\Sigma_\psi).
\end{align}

The flow equation for the contact becomes
\begin{align}
 &\nonumber \partial_t C_k =\frac{h^2_\phi}{4}(\eta_A \rho_{0,k}+\partial_t \rho_{0,k})\\
 \label{B-15c} &+\frac{h_\phi^2}{4} \int_{\vec{q}}\frac{\dot{R}_\phi}{A} T \sum_n \Bigl(\frac{1}{\mbox{det}_B(Q)}-\frac{2S^2\omega_{\phi,k}^2+2\lambda^2\rho_0^2}{\mbox{det}^2_B(Q)}\Bigr).
\end{align}
The result of the Matsubara summation and the $\vec{q}$--integration is given in Eq. (\ref{3-29}).

To clarify the relation between $\partial_tC_k$ and $\partial_t n_{\phi,k}$, we consider the scale-dependent density $n_k$ defined by
\begin{align}
\label{B-15} n_k = -\frac{\partial U_k}{\partial \mu}(\rho_{0,k}).
\end{align}
For $k=0$ we arrive at the physical density 
\begin{align}
 n =\frac{\partial P(\mu,T)}{\partial \mu} = - \frac{\partial U(\mu,T,\rho_0)}{\partial \mu}
\end{align}
with pressure $P(\mu,T)$. The corresponding flow equation is given by
\begin{align}
 \label{B-16} \partial_t n_k = -\alpha_k (\eta_A \rho_{0,k}+\partial_t\rho_{0,k}) -\Bigl(\partial_\mu \partial_t|_{\bar{\rho}}U_k\Bigr)(\rho_{0,k}).
\end{align}
with
\begin{align}
 \label{B-17} \alpha_k = \frac{\partial^2 U}{\partial\mu\partial\rho}(\rho_{0,k}).
\end{align}
The flow of the effective potential receives contributions from bosonic and fermionic fluctuations. Defining
\begin{align}
 \label{B-18} \partial_t n_{\phi,k}^{(U)} = -\Bigl(\partial_\mu \partial_t|_{\bar{\rho}}U^{(B)}_k\Bigr)(\rho_{0,k}),\\
 \label{B-18b} \partial_t n_{\psi,k}^{(U)} = -\Bigl(\partial_\mu \partial_t|_{\bar{\rho}}U^{(F)}_k\Bigr)(\rho_{0,k})
\end{align}
as the bosonic or fermionic contribution, respectively, we can employ
\begin{align}
 \nonumber &\Bigl(\partial_t|_{\bar{\rho}}U_k^{(B)}\Bigr)(\rho) \\
 \label{B-19} &= \int_Q \frac{\dot{R}_\phi}{A} \frac{k^2+U_k'(\rho)+\rho U_k''(\rho)}{[k^2+U'(\rho)+2\rho U_k''(\rho)][k^2+U'(\rho)]+S^2q_0^2}
\end{align}
to find
\begin{align}
 \label{B-20} \partial_t n_{\phi,k}^{(U)} = -\alpha_k \int_Q \frac{\dot{R}_\phi}{A} \Bigl(\frac{1}{\mbox{det}_B(Q)}-\frac{2S^2\omega_{\phi,k}^2+2\lambda^2\rho_0^2}{\mbox{det}_B^2(Q)}\Bigr).
\end{align}
This is precisely the third term in the flow equation  (\ref{B-14}) for $ C_k$. We can summarize these findings in the generally valid relation
\begin{align}
\nonumber \partial_tC_k &= -\frac{h^2_\phi}{4\alpha_k} \Bigl( -\alpha_k(\eta_A \rho_{0,k}+\partial_t \rho_{0,k}) + \partial_t n_{\phi,k}^{(U)}\Bigr)\\
\label{B-21}&= -\frac{h^2_\phi}{4\alpha_k}  \Bigl(\partial_t n_k - \partial_t n_{\psi,k}^{(U)}\Bigr),
\end{align}
where $n^{(U)}_{\phi/\psi,k}$ are contributions which arise from effects of bosonic/fermionic fluctuations on the effective potential and hence the density. In contrast, the term $-\alpha_k(\eta_A \rho_{0,k}+\partial_t \rho_{0,k})$ accounts for the nontrivial scaling of the renormalized propagator and the contribution from condensed bosons.

In the BEC limit, Eq. (\ref{B-21}) simplifies considerably. Indeed, following the flow of $C_k$ from $k=\Lambda$ to $k=0$ we see that in the early stages of the flow, where $k$ is much larger than the many body-scales set by $\mu$ and $T$, the flow of $C_k$ is zero, because $\rho_{0,k}=0$ and there are no bosonic fluctuations on high energy scales. Hence
\begin{align}
\label{B-22} \partial_k C_k =0\ \text{ for } k\gg \mu^{1/2}, T^{1/2}.
\end{align}
However, the flow of the prefactor $h^2_\phi/\alpha_k=\bar{h}^2_\phi/\bar{\alpha}_k$ is governed by the scale set by the scattering length $a^{-1}$. On the far BEC side, this quantity is large and these renormalization effects set in far above the many-body scales. At such high scales the vacuum relation $\bar{\alpha}_k = - 2 Z_{\phi,k}$, with wave function renormalization $Z_{\phi,k}$ of the bosons, is valid. The relation stems from the appearance of the combination $Z_{\phi,k}(\partial_\tau-2\mu)$ in the propagator due to semi-local ${\rm U}(1)$--invariance and the symmetry preserving nature of the flow equation. One can show that the vacuum flow of $Z_{\phi,k}$ is solved by $Z_{\phi,k=0}=\bar{h}^2_\phi a/32\pi$ on the BEC side \cite{PhysRevA.76.021602}, i.e.
\begin{align}
 \label{B-23} -\frac{\bar{h}^2_\phi}{4\bar{\alpha}_k} \to \frac{4\pi}{a} \text{ for } k\gg \mu^{1/2}, T^{1/2}.
\end{align}
This prefactor effectively enters the renormalization group equation of $C_k$ which takes place on the many-body scales and thus we can write
\begin{align}
 \label{B-24} \partial_k C_k = \frac{4\pi}{a} \partial_k n_{\phi,k}.
\end{align}
Due to the fact that there are no fermion fluctuations contributing to the density on the BEC side, we have $n_{\phi,k} = n_k$ and thus arrive at the well-known relation $C_{\rm BEC} = 4\pi n/a$. The mean-field result $C_{\rm BEC}=4\pi n/a$ receives corrections from bosonic fluctuations, which are incorporated in the renormalization group flow.

\section{Perturbative contact from the equation of state}
\label{AppC}
We consider the perturbative BEC regime of the crossover, i.e. the region
\begin{align}
\label{C-1} (k_{\rm F}a)^{-1} \gg 1.
\end{align}
Moreover, we restrict the considerations to the case of zero temperature. We  closely follow Tan's presentation in Ref. \cite{Tan20082971}. The energy density of the system (in the canonical variables) is given by the Lee--Huang--Yang expression
\begin{align}
 \label{C-2} \frac{E}{V} = n_{\rm d} \vare_{\rm B} + \frac{g_{\rm d}}{2} n_{\rm d}^2 \Bigl(1+\frac{128}{15\sqrt{\pi}}\sqrt{n_{\rm d}a_{\rm d}^3}\Bigr),
\end{align}
where $n_{\rm d}$ is the number of dimer atoms, which is related to the density according to $n_{\rm d}=n/2$ on the far BEC side, and $\vare_{\rm B}$ is the binding energy of a dimer. From Eq. (\ref{C-1}) we find the gas parameter $n_{\rm d}a_{\rm d}^3$ to be small. The coupling constant for the dimers is given by
\begin{align}
 \label{C-3} g_{\rm d} = \frac{4\pi a_{\rm d}}{M_{\rm d}} = 4\pi \kappa a,
\end{align}
where we used that $M_{\rm d}=2M=1$ and $a_{\rm d}=\kappa a$. The dimensionless constant $\kappa$ relates the scattering length of dimers to the fermionic scattering length. The exact value of $\kappa$ is known to be $0.6$ \cite{PhysRevLett.93.090404}. However, within our truncation we have $\kappa=0.72$ from a solution of the vacuum problem \cite{ANDP:ANDP201010458}. We employ the latter value for consistency and emphasize that $\kappa$ is not a free parameter in our model. The energy density in terms of $n$ and $a$ is found to be
\begin{align}
 \label{C-4} \frac{E}{V} = n \frac{\vare_{\rm B}}{2} + \frac{\pi \kappa a n^2}{2} \Bigl(1+\frac{128}{15\sqrt{2\pi}}\sqrt{n \kappa^3a^3}\Bigr).
\end{align}
We compute the chemical potential according to $\mu(n)=\mbox{d}(E/V)/\mbox{d}n$ and subtract half the binding energy to obtain the (positive) many-body chemical potential
\begin{align}
 \label{C-5} \mu_{\rm mb}(n) = \mu - \frac{\vare_{\rm B}}{2} = \pi \kappa a n\Bigl(1+\frac{32}{3\sqrt{2\pi}}\sqrt{n\kappa^3a^3}\Bigr).
\end{align}
Inverting this relation to the same order of approximation we find
\begin{align}
 \label{C-6} n(\mu_{\rm mb}) = \frac{\mu_{\rm mb}}{\pi \kappa a}\Bigl(1-\frac{32\kappa}{3\pi\sqrt{2}}\sqrt{\mu_{\rm mb}}a\Bigr).
\end{align}
This is the equation of state in the grand canonical variables at $T=0$ to leading order in the gas parameter $\sqrt{\mu_{\rm mb}}a$.

By virtue of the adiabatic sweep theorem (\ref{1-2}) we can derive an expression for the contact in the BEC regime. For this purpose we employ $\vare_{\rm B}=-1/Ma^2=-2/a^2$. From Eq. (\ref{C-4}) we then find
\begin{align}
 \label{C-7} C_{\rm BEC} = \frac{4\pi n}{a} + \kappa \pi^2 a^2n^2 \Bigl(1+\frac{64}{3\sqrt{2\pi}}\sqrt{n\kappa^3a^3}\Bigr)
\end{align}
at zero temperature. By dividing by $k_{\rm F}^4$ we obtain   
\begin{align}
 \nonumber \frac{C_{\rm BEC}}{k_{\rm F}^4} = \mbox{ }&\frac{4}{3\pi}(k_{\rm F}a)^{-1}\\
 \label{C-8} &+\frac{\kappa}{9\pi^2}(k_{\rm F}a)^2\Bigl(1+\frac{64 \kappa^{3/2}}{3\pi\sqrt{6\pi}}(k_{\rm F}a)^{3/2}\Bigr).
\end{align}
Inserting the equation of state (\ref{C-6}) we arrive at
\begin{align}
 \label{C-9} \frac{C_{\rm BEC}}{\mu_{\rm mb}^2} = \frac{4}{\kappa}(\sqrt{\mu_{\rm mb}}a)^{-2}\Bigl(1-\frac{32\kappa}{3\pi\sqrt{2}}\sqrt{\mu_{\rm mb}}a\Bigr).
\end{align}

We now consider the asymptotic behavior on the BCS side. The equation of state at zero temperature is given by
\begin{align}
 \label{C-10} \frac{E}{V} = \frac{3}{5}k_{\rm F}^2 n \Bigl(1+\frac{10}{9\pi}k_{\rm F}a\Bigr).
\end{align}
We deduce for the chemical potential that
\begin{align}
 \label{C-11} \mu(k_{\rm F}) = \frac{\mbox{d}}{\mbox{d}n} \frac{E}{V} = k_{\rm F}^2\Bigl(1 + \frac{4}{3\pi}k_{\rm F} a\Bigr),
\end{align}
The contact is found from
\begin{align}
  \label{C-13}C &= 2 \pi \frac{\mbox{d}(E/V)}{\mbox{d}(-1/a)} =4\pi^2 n^2 a^2.
\end{align}
Thus we have
\begin{align}
  \label{C-14}\frac{C_{\rm BCS}}{k_{\rm F}^4} = \frac{4}{9\pi^2} (k_{\rm F}a)^{2}.
\end{align}
Inserting the equation of state $k_{\rm F}(\mu) = \sqrt{\mu}$ we then arrive at
\begin{align}
  \label{C-16}\frac{C_{\rm BCS}}{\mu^2} &= \frac{4}{9\pi^2}(\sqrt{\mu}a)^2.
\end{align}

\vfill

\end{appendix}


\eject

\bibliographystyle{apsrev4-1}
\bibliography{referencesTAN}

\end{document}